\documentclass{aastex62}

\received{..., 2018}
\revised{...., 2018}
\accepted{\today}
\submitjournal{ApJ}

\shorttitle{Effects of continuum opacity fudging on synthesis of stellar spectra}
\shortauthors{Criscuoli}

\begin{document}

\title{Effects of continuum fudging on non-LTE synthesis of stellar spectra. \\I. Effects on estimates of UV continua and Solar Spectral Irradiance variability.}

\correspondingauthor{Serena Criscuoli}
\email{scriscuo@nso.edu}

\author[0000-0002-4525-9038]{Serena Criscuoli}
\affil{National Solar Observatory\\
3665 Discovery Dr. \\
Boulder, CO 80303, USA}



\begin{abstract}
Syntheses performed under non Local Thermodynamic Equilibrium (non-LTE) conditions usually overestimate stellar spectra.  An approach widely adopted in the literature to reduce the excess of UV radiation consists of artificially increasing the continuum opacity by using multiplicative fudge factors, which are  empirically derived to impose the synthetized spectrum to match the observed one.  Although the method was  initially developed to improve non-LTE synthesis of spectral lines, it has been recently employed to model solar spectral irradiance variability. Such irradiance  reconstruction techniques combine spectral synthesis of different types of structures, which are performed making use of factors derived from a reference, quiet Sun model. Because the opacity scales in a complex way with plasma physical properties, the question arises whether, and to what extent, fudge factors derived using a reference model can be used to adjust the opacity of models representing different types of quiet and magnetic features. Here we investigate the effects of  opacity fudging on estimates of solar and stellar irradiance variability in UV bands. We find that the use of fudge factors might underestimate the variability by 19\% and up to 20\%
in the ranges 230-300 nm and 300-400 nm, respectively. These estimates are model dependent and should be considered as upper limits. Finally, our analysis suggests the uncertainties generated by the use of fudge factors to increase with the decrease of stellar metallicity and to be significant for stars
whose variability is facula-dominated and whose effective temperature is larger than approximately 4000 K.

\end{abstract}



\section{Introduction} \label{sec:intro}
Synthesis of stellar spectra require proper modeling of the physical processes that describe the interaction between stellar plasma and radiation. These, in turn, require 
detailed knowledge of atomic and molecular parameters.
Especially in the UV syntheses are affected by uncertainties in atomic photo-ionization parameters \citep[e.g.][]{bell1991}, missing line opacities \citep[e.g.][]{allendeprieto2003, short2009} 
and adopted atomic
models \citep[e.g.][]{mashonkina2011}.
As a result, the UV opacity is underestimated. The problem is particularly evident in non-LTE synthesis, as missing UV opacity leads to over-ionization of metals, especially Iron, 
and consequent weakening of line strengths. Therefore, non-LTE computations of 
stellar spectra generally provide larger overestimate of stellar radiative fluxes than LTE synthesis \citep[e.g.][]{allendeprieto2003,fontenla2009,shapiro2010, youngshort2017}. 
This issue is often addressed in the literature as the problem of `missing UV opacity'.  Modeling efforts by \citet{fontenla2015} using the Solar Radiation Physical Modeling system (SRPM),
 concluded that most of the missing opacity should be ascribed to molecular photo-dissociation processes, especially NH, not properly taken into account in most of the available radiative transfer codes. 

The problem of `missing opacity' is extremely important for proper non-LTE synthesis of lines, especially of those for which the radiation field plays an important role in determining the ionization balance, such as Fe~I, Mg~I, Ni~I and Ca~I 
\citep[e.g.][]{bruls1992,langangen2009, busa2001, collet2005}. 
The issue is complicated by the fact that synthetizing the  millions of lines that populate the UV and modeling  hundreds of photo-ionization transitions is computationally demanding, and still prohibitive in the case
of synthesis performed using three-dimensional magneto hydrodynamic (3D MHD) simulations
of stellar atmospheres \citep[e.g.][]{nordlander2017}.  In general, when explicit  computation is not feasible, background opacities are taken into account using pre-computed 
opacity tables \citep[e.g.][]{zhao1998,falchi1998, andretta2005, amarsi2016} or ad-hoc ``fudge factors'' \citep[e.g.][]{bruls1992,busa2001,trujillobueno2009, shapiro2011, sasso2017}. The latter is the subject of this paper,
and it consists in adjusting (fudging) the continuum opacity with wavelength dependent multiplicative factors derived imposing the synthetic spectrum to match the observed one.      

Although the method was developed mostly for the non-LTE synthesis of
spectroscopic lines, it has been recently employed to synthetize spectra used in semi-empirical reconstructions of solar spectral irradiance variability \citep[e.g.][]{shapiro2010, shapiro2011b, thuillier2014, criscuoli2018}. These reconstructions are based on the assumption that irradiance variability is modulated, at least on the solar cycle temporal scale, by the presence of magnetic features over the solar surface. Irradiance is thus reconstructed by weighing synthetic spectra of quiet and magnetic features with the corresponding area coverage, typically derived by daily full-disk observations of the Sun. In these reconstructions, the same fudge factors 
derived using a quiet Sun model are employed for the non-LTE synthesis of spectra of magnetic regions (network, faculae, sunspots). This is motivated by the lack of systematic multi-wavelength measurements of fluxes of different types of magnetic structures, which would be necessary to compute the multiplicative factors for the different corresponding atmosphere models. Because stellar opacities show complicated dependence with the physical properties of stellar atmospheres, we expect this approach to introduce uncertainties in synthetic spectra of magnetic structures and therefore in irradiance reconstructions. For completeness, it should be noted that not all spectral irradiance reconstructions making use of non-LTE synthesis  employ fudge factors. In particular, fudge factors are employed in reconstructions that use the codes COSI \citep{haberreiter2008, shapiro2010} and its evolution NESSY \citep{tagirov2017}, and RH \citep{uitenbroek2001}. Reconstructions performed using the SRPM  system \citep[e.g.][]{fontenla2011,fontenla2015} do not make use of fudging. We refer the interested reader to the recent reviews on irradiance reconstructions techniques by \citet{ermolli2013} and \citet{yeo2014}. 

Studies aiming at measuring and modeling solar radiation and its variability are stronlgy motivated by the impact that solar \textbf{irradiance (that is, the electromagnetic energy emitted by the Sun  received at the top of the Earth atmosphere in the unit of area and time)}, especially in the UV, has on 
the chemistry and physical properties of the Earth atmosphere and climate \citep[e.g.][]{gray2010,matthes2017}. 
Studies of solar variability have been recently also driven by the necessity of improving our understanding of stellar variability \citep[see][for a recent review]{fabbian2017}, 
which, in turn, is essential to characterize the habitable zones of stars and the atmospheres of their exoplanets. As for the Earth, modeling of exoplanet atmospheres requires as 
fundamental input the spectral energy distribution of the hosting star, especially UV and shorter wavelengths \citep[e.g.][]{tian2014,ranjan2017,
rugheimer2018}. Unfortunately, measurements of UV radiation are strongly hampered by the interstellar medium absorption (up to 70-90\%), which is \textbf{significant}
even for relatively close stars,
so that estimates of stellar UV radiation stongly relies on modeling \citep[see][for a recent review]{linsky2017}.
Moreover, because there is no mission scheduled in the next future to observe stellar spectra in the UV, after the  Hubble Space Telescope will cease operations the 
characterization of UV spectra of stars hosting exoplanets that will be discovered by current and future missions (e.g. TESS or James Webb Space Telescope)
will necessarely rely on indirect estimates, performed for instance through the use of semi-empirical models \citep[e.g.][]{mauas1997,fontenla2016, busa2017} or proxies 
\citep[e.g.][]{stelzer2013,shkolnik2014}. Stellar irradiance variability also affects the detectability of exoplanets \citep[see for instance the recent review by][]{oshagh2018}. The passage over the disk of spots and faculae may induce photometric variations  
of similar or larger amplitude than photometric variations induced by planetary transits. 
Moreover, the presence of active regions may alter spectral line profiles, thus hindering exoplanet detections performed through radial-velocity
measurements. Similarly, spectroscopic techniques that allow to estimate the physical properties of exoplanet atmospheres 
\citep[see e.g.][for a recent view]{kreidberg2017} require as fundamental input the spectra synthetized through models representing quiet and active regions (faculae and sunspots). Finally, stellar irradiance variability observed at different spectral ranges, especially in the UV, 
is a fundamental observable for the characterization of the magnetic activity of a star, and therefore for the understanding of dynamo processes in stellar 
objects \citep[e.g.][]{reinhold2013, salabert2016, basri2016}. 
Because stellar photometric and spectral variability can  be modeled using the semi-empirical approaches developed for the Sun described 
above \citep[see for instance][]{shapiro2016, witzke2018},    
understanding the limitations of current irradiance models is fundamental to improve our 
capability of modeling stellar variability.  

The purpose of this paper is to investigate the effects of using fudge factors derived for quiet Sun atmosphere models for the computation of spectra of different atmospheric structures (e.g. network, dark lanes, sunspots, etc.).
We focus here on the effects of fudging on estimates of UV continuum contrast of magnetic structures and of UV irradiance variability.  In a paper in preparation we will focus on the effects of fudging the continuum opacity on the non-LTE synthesis of spectral lines
sensitive to the UV background radiation. 

The paper is structured as follows: in Sec.~\ref{sec:synthe} we describe the details of the spectral synthesis and methodology employed to investigate the use of fudge factors; 
in Sec.~\ref{sec:results} we present the differences obtained between UV spectra synthetized using a full-set of available opacities as opposed to the use of fudge factors; 
in Sec.~\ref{sec:variab} we investigate the effects of such differences on computations of magnetic structures contrast and estimates of Spectral Irradiance variability.
The discussion of our results in the framework of solar and stellar irradiance studies and our conclusions are presented in Sec.~\ref{sec:disc}.

\section{Spectral synthesis} \label{sec:synthe}
To estimate the effects of fudging on synthetic spectra, we proceeded as follows. We first performed two syntheses using a quiet Sun model,  differing for the atomic and molecular inputs employed.
The first one was taken as the ``reference spectrum''. The second was performed with atomic and molecular inputs as to produce, without fudging,  a flux larger than the reference one. Fudge factors were then estimated by minimizing the difference between the two spectra, both convolved with a Gaussian function 1 nm wide. Spectral syntheses of magnetic features were then computed using the same two sets of atomic
and molecular inputs. Spectra obtained using the input same parameters employed to compute the ``reference'' spectrum were considered the ``true'' spectra, while 
the second set of spectra was computed  
 \textit{using the fudge factors determined for the quiet Sun model}. Spectra obtained with the two methods were then finally compared. Because different parts of the spectrum are affected differently by the presence of spectral lines 
 and photolysis processes, we used different atomic 
and molecular inputs for different spectral regions, as explained in detail below. 

\begin{figure*}
\centering
\includegraphics[width=8.5cm]{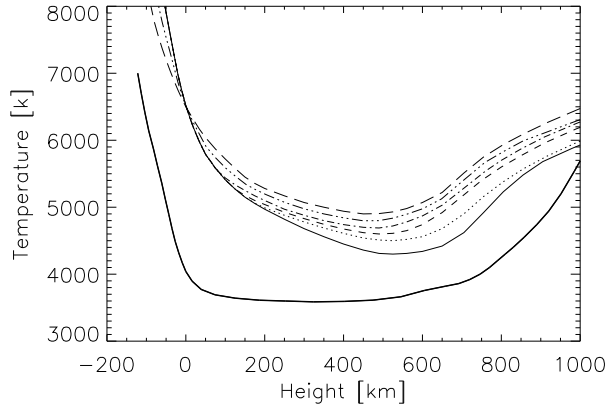}
\caption{\label{Fig_temps} Temperature stratification of the FAL99 models in the photosphere and lower chromosphere.   Continuous: model-A. Dotted: model-C. Dashed: model-E.
Dot-dashed: model-F. Dot-dot-dot-dashed: model-H. Long-dashed: model-P. Thick line: model-S.}
\end{figure*}

For the synthesis we employed the RH code \citep{uitenbroek2001}, which  allows to solve the radiative transfer problem in non-LTE. RH also allows to compute the background line-opacities 
by LTE synthesis of sets of molecular and atomic lines, and/or by applying fudge factors to the continuum opacity. 
All the synthesis were performed using one-dimensional, static atmosphere models of quiet and magnetic features described in \citet{fontenla1999} (FAL99, hereafter).
These consist of a set of models representing: the average quiet Sun (model-C),
a cold supergranule center (model-A),  quiet (model-E) and bright (model-F) network, facula (model-H) and bright-facula (model-P), and, finally, sunspot umbra (model-S).
Note that syntheses obtained with model-C were employed to derive the fudge factors. The temperature stratification of the various atmosphere models are illustrated in Fig.~\ref{Fig_temps}. 
These sets of models were chosen to facilitate the comparison of our results with those reported in \citet{shapiro2010}.  Finally, for all our calculations we employed 
the abundance values published in \citet{grevesse1991}.
 
\section{Effects of fudging on UV spectra of magnetic structures} \label{sec:results}

Different parts of the spectrum are  differently affected by atomic and molecular inputs. We therefore performed synthesis for the 230-300 nm and for the 300-400 nm ranges separately. 

\subsection{Spectral range [230-300] nm}  \label{sec:s200_300}

The contribution of photo-ionization of metals to the continuum opacity increases with the decrease of the wavelength, to become the major contribution at wavelengths shorter than approximately 280 nm \citep{allendeprieto2003, fontenla2011}. 
In the range [200-300] nm, the major contributors are Fe~I and Mg~I. Recently, \citet{mashonkina2011} showed that a correct estimate of the Fe I contribution requires the use of a model atom that includes more than 3000 levels. Because most 
of the non-LTE codes 
typically do not employ such complete Fe atoms \citep[for instance  ][ employed an atom with a few hundred levels]{fontenla2015}, it is reasonable to assume that the incomplete description of the Fe I transitions and photo-ionizations might be one of the major source of discrepancy between non-LTE synthetic spectra and observations.  We therefore performed 
two sets of synthesis. In the first one (the reference) the Fe was synthetized in LTE, while in the second one  Fe was computed in non-LTE and fudge factors where applied to match the reference one.  In both cases we employed a 52-levels 
Fe~I model atom, updated with the most recent photo-ionization cross-sections computed by \citet{bautista2017}. In both synthesis H, He, C, Ca, Al, Si and Mg where computed in non-LTE. Photo-dissociation and bound-bound transitions for the most abundant diatomic species (e.g. AlO, CH, CN, CO, H2, MgH, NH, TiO, etc.), 
as well as 
atomic bound-bound transitions from the Kurucz database \footnote{available at  http://kurucz.harvard.edu/} were also included in both syntheses. Note that  RH 
automatically excludes  from the Kurucz line-list all bound-bound transitions
specified in the model atoms. As explained above, the fudge factors were computed imposing the flux synthetized for the quiet Sun model (model-C) with Fe~I in non-LTE to match the spectrum computed with Fe~I in LTE.  Comparison of the spectra obtained from the reference synthesis, from the synthesis performed using the second set of opacities without fudging the continuum opacity, and the observed solar spectrum is provided in Fig.~\ref{referefluxes}.

\begin{figure*}
\centering
\includegraphics[width=8.5cm]{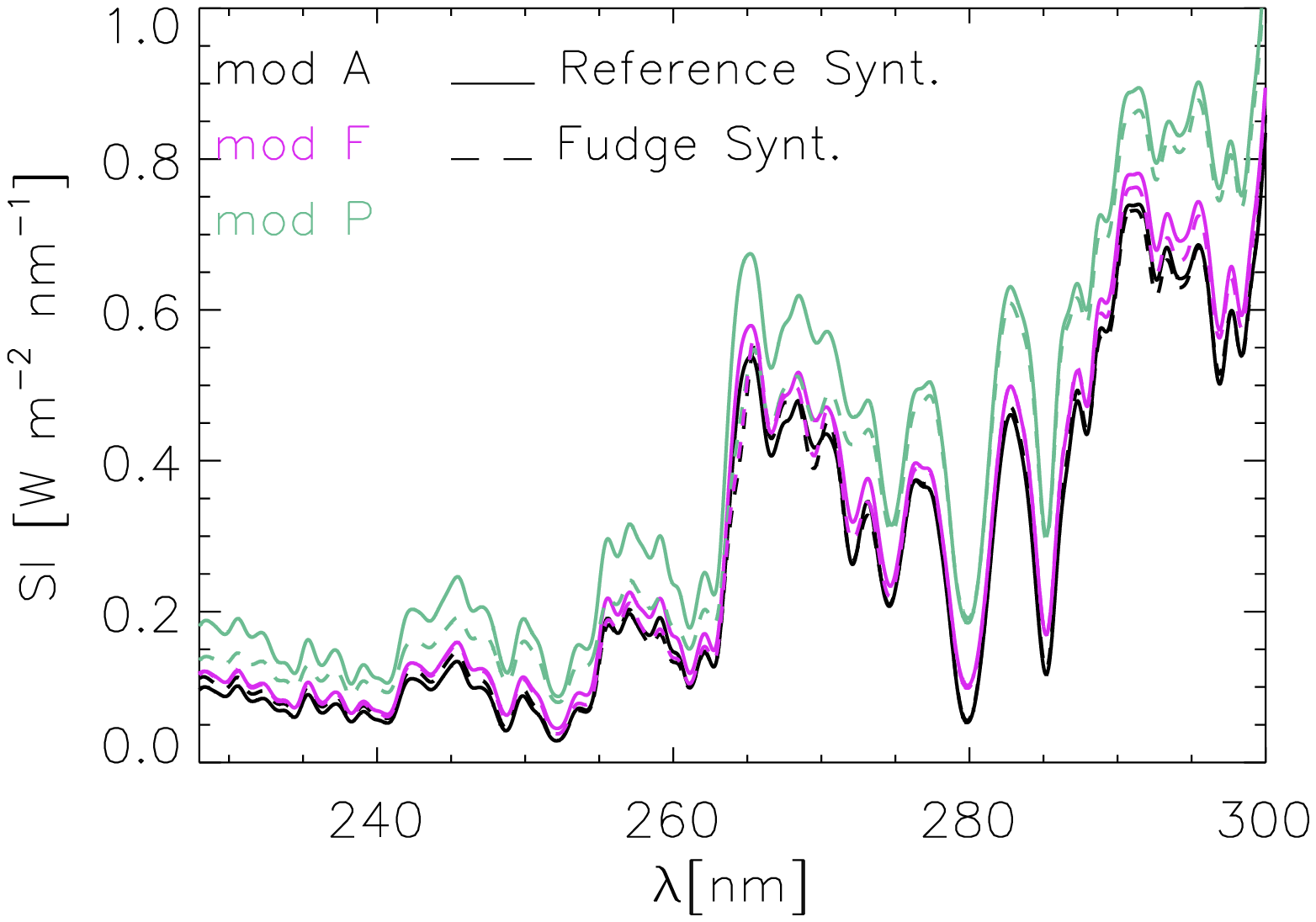} \includegraphics[width=8.5cm]{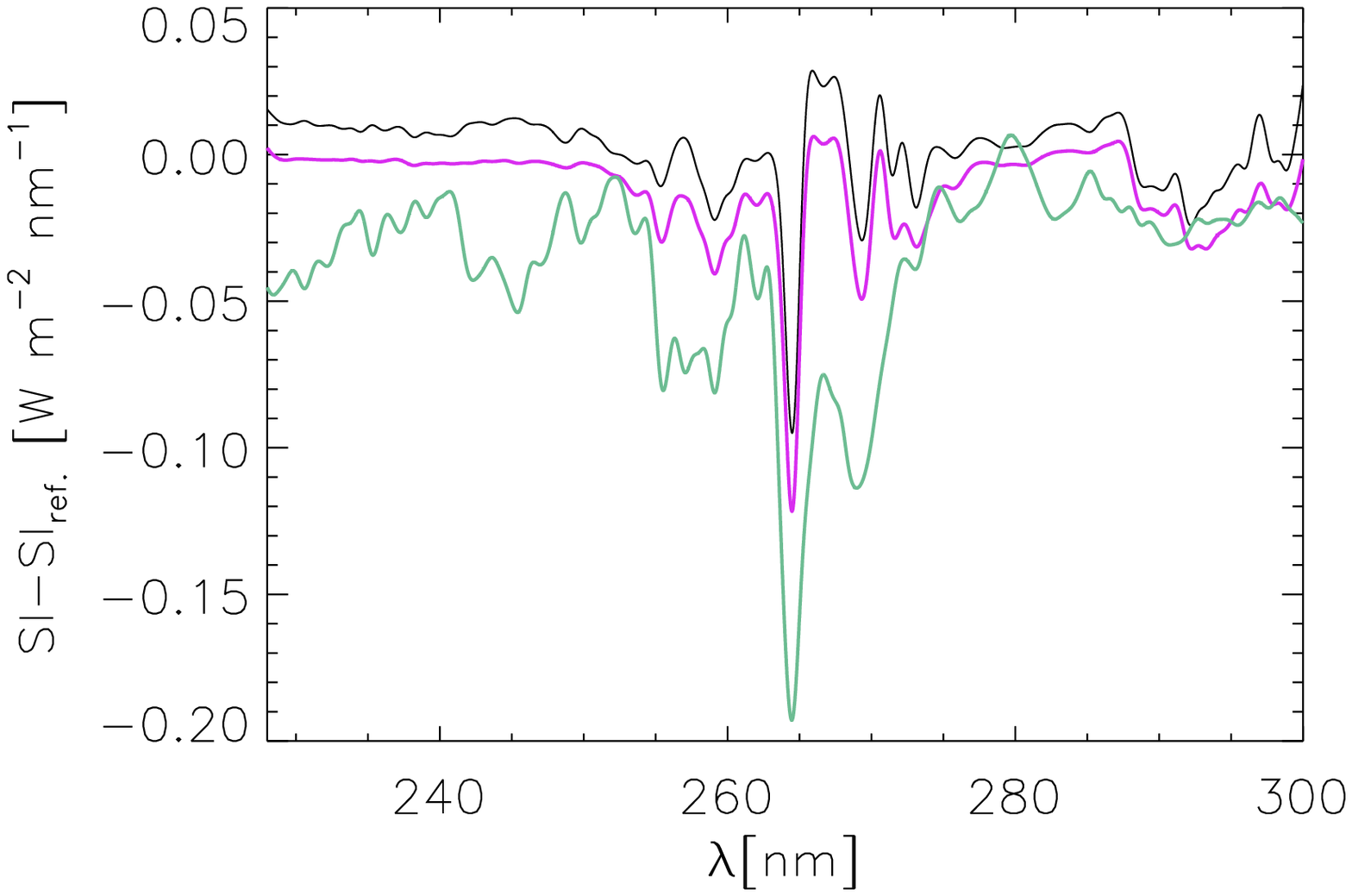}\\
\includegraphics[width=8.5cm]{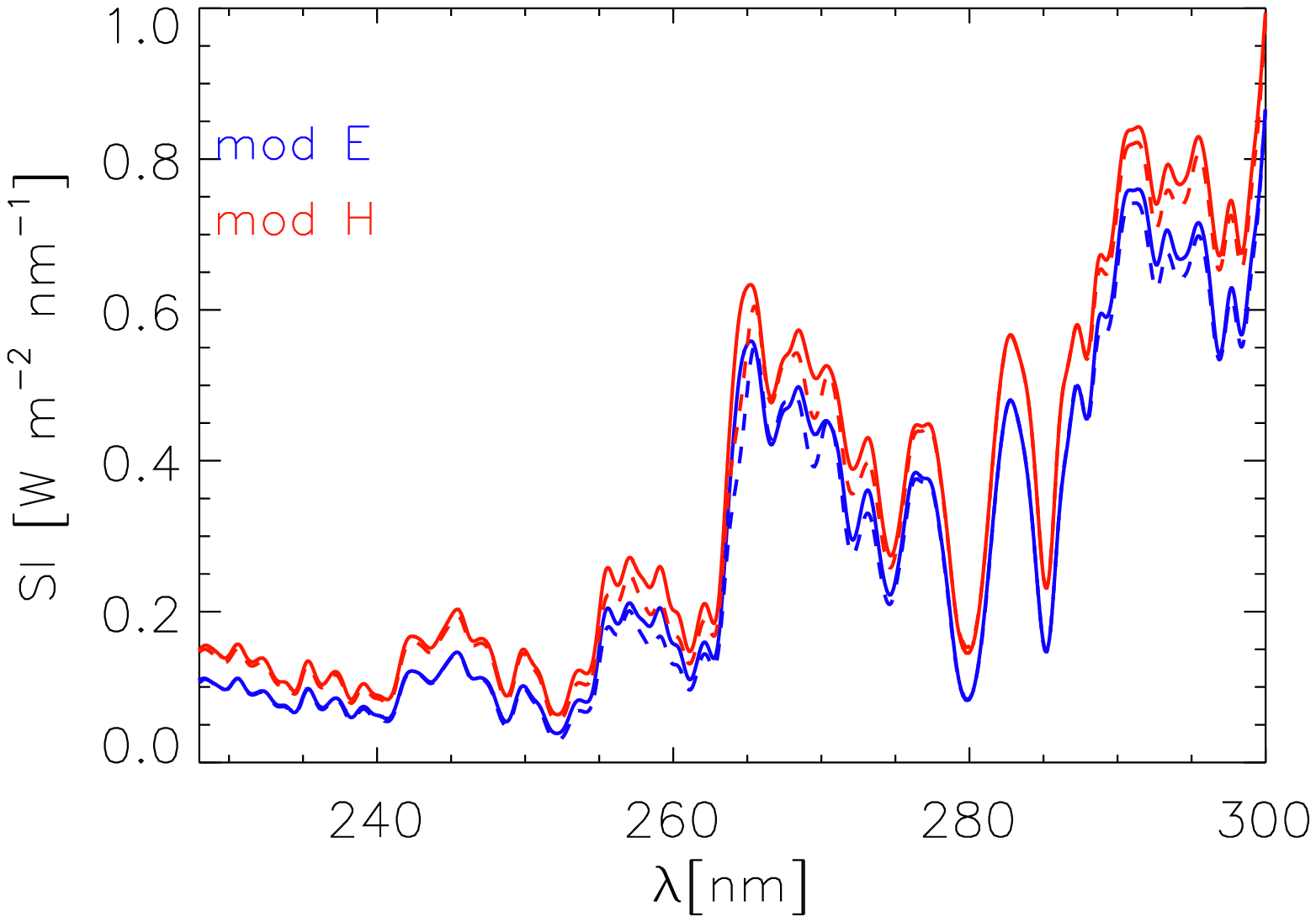} \includegraphics[width=8.5cm]{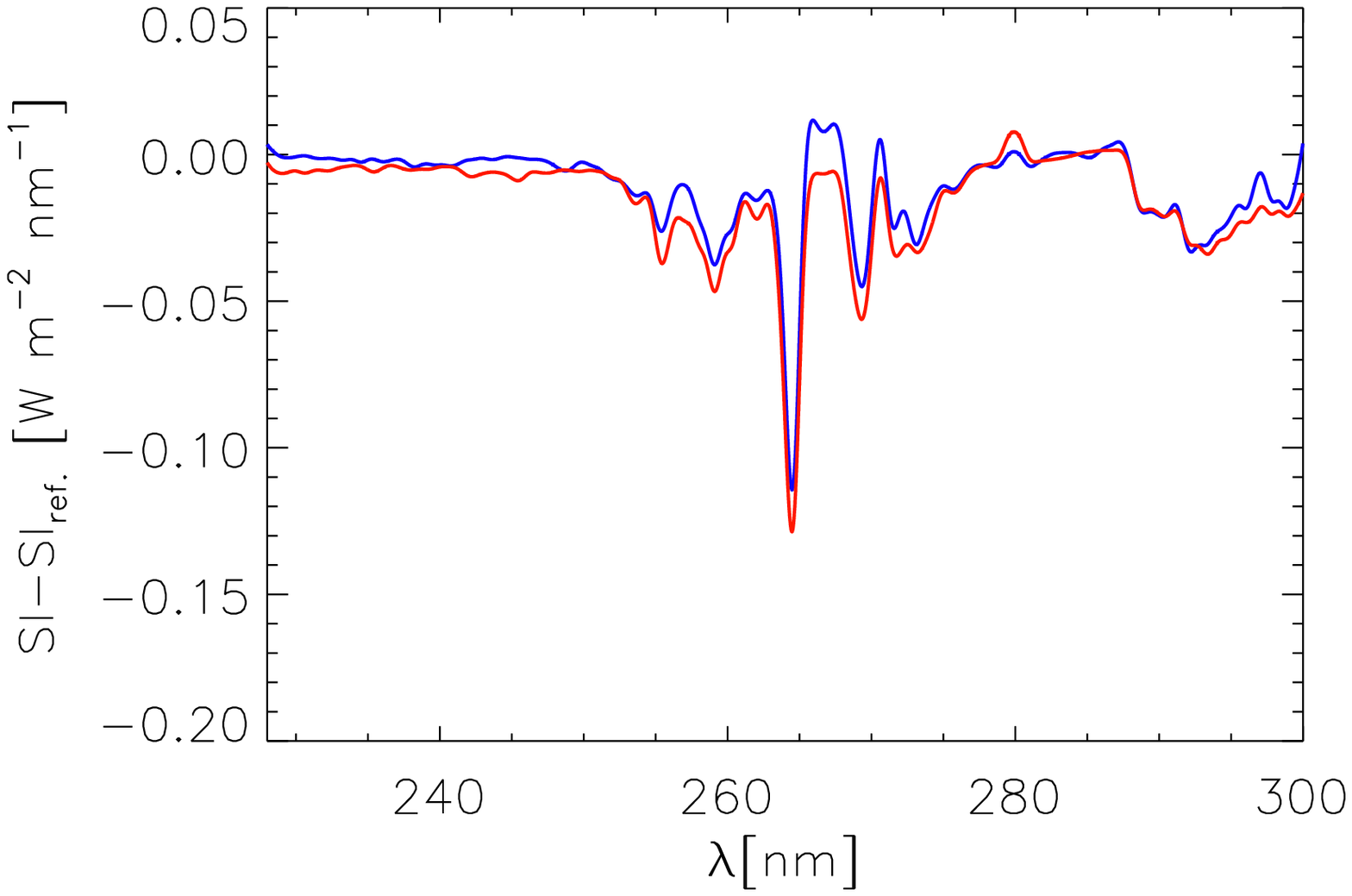}\\
\includegraphics[width=8.5cm]{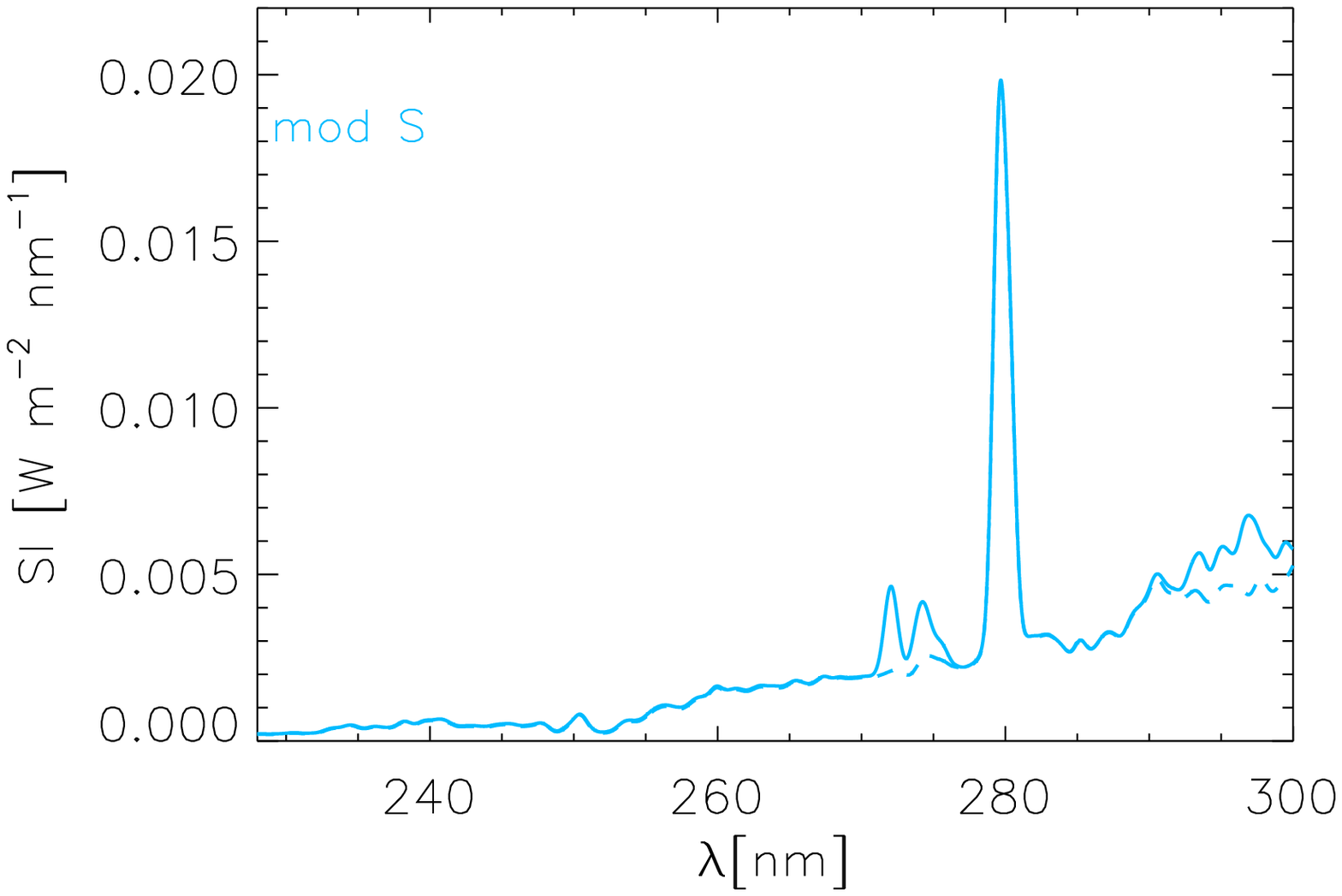} \includegraphics[width=8.5cm]{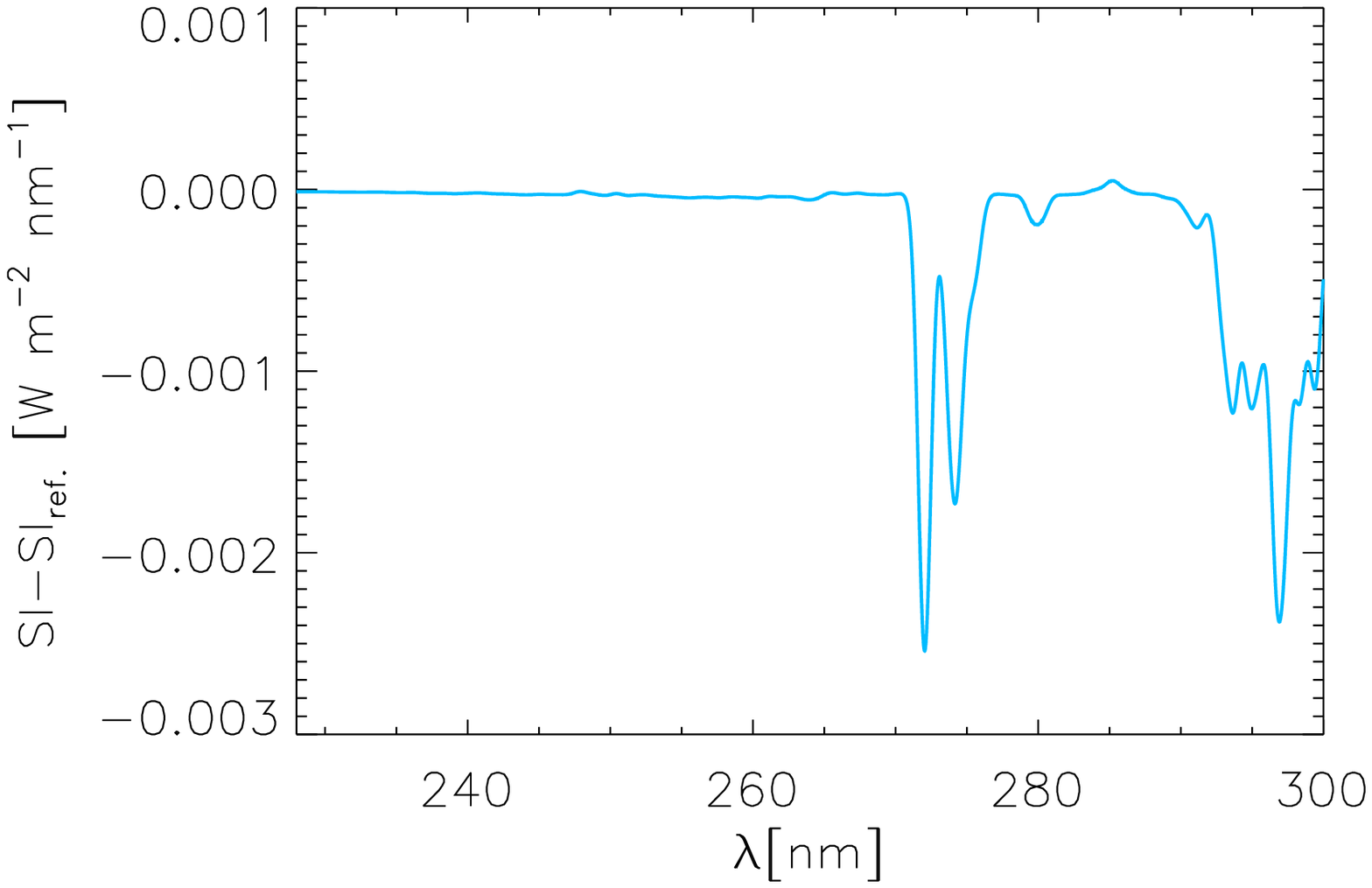}\\
\caption{\label{Fig_200_300} Left: Spectral Irradiance obtained for different atmosphere models from the reference synthesis (continuum line) and applying fudge factors to the continuum opacity (dashed line). Right: difference between the two SI synthesis obtained 
for different atmosphere models.
Synthetic spectra were convolved with a 1 nm wide Gaussian function. }
\end{figure*}

Figure~\ref{Fig_200_300} shows the Spectral Irradiance (SI) obtained for various atmosphere models using the two sets of atomic inputs (left) and their differences (right). The plots show that in general the difference between the reference synthesis
and the synthesis
obtained using fudge factors increases with the increase of the difference between the temperature stratification of the model and model-C (c.f.r. Fig.\ref{Fig_temps}). In particular, for model-A the irradiance is overall overestimated while for the other models the irradiance is overall underestimated.
Note that, instead, for the sunspot model-S differences are one order of magnitudes smaller than for the other models. 
These results are discussed more in detail in Sec.~\ref{sec:disc}.

\subsection{Spectral range [300-400] nm} \label{sec:s300_400}
The importance of the inclusion of molecular lines for the computation of the solar spectrum and solar irradiance variability, especially in the UV,  has been outlined in various papers \citep[e.g.][]{fontenla2015, shapiro2015}. 
In the UV, the range between 300 and 400 nm is dominated by CN, OH and NH bands, while the contribution to the continuum opacity of metals is less important \citep[e.g.][]{bruls1992, castelli2004}. For this spectral range we therefore 
investigated the case in which fudging is necessary mostly because of ``missing'' molecular lines.  Following the procedure explained above, we  evaluated the emergent flux for various atmosphere models using two sets of inputs, 
in the first one (the reference synthesis) the full  molecular list from the Kurucz database was used, in the second one, molecular lines and photo-dissociation were not taken into account, and the synthesis was performed fudging the continuum opacity. As above, the fudging coefficients 
as function of wavelength were derived imposing the model-C flux computed without the inclusion of molecules to match the model-C flux computed including molecules. In both cases, H, He, C, CA, 
Al, Fe, Si and Mg where computed in non-LTE. Comparison of the spectra obtained from the reference synthesis, from the synthesis performed using the second set of opacities without fudging the continuum opacity, and the observed solar spectrum is provided in Fig.~\ref{referefluxes}.

\begin{figure}
\centering
\centering
\includegraphics[width=8.5cm]{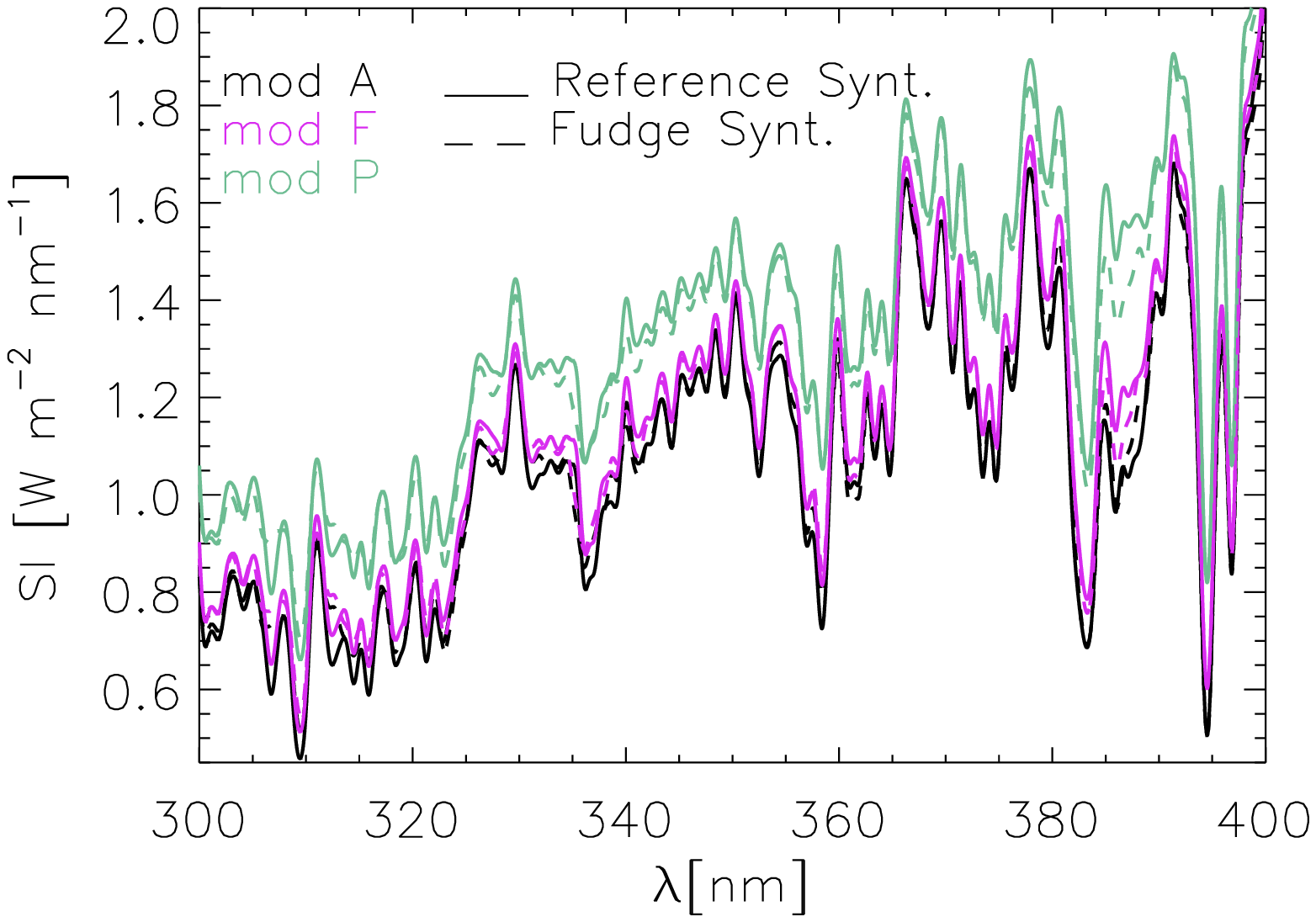}\includegraphics[width=8.5cm]{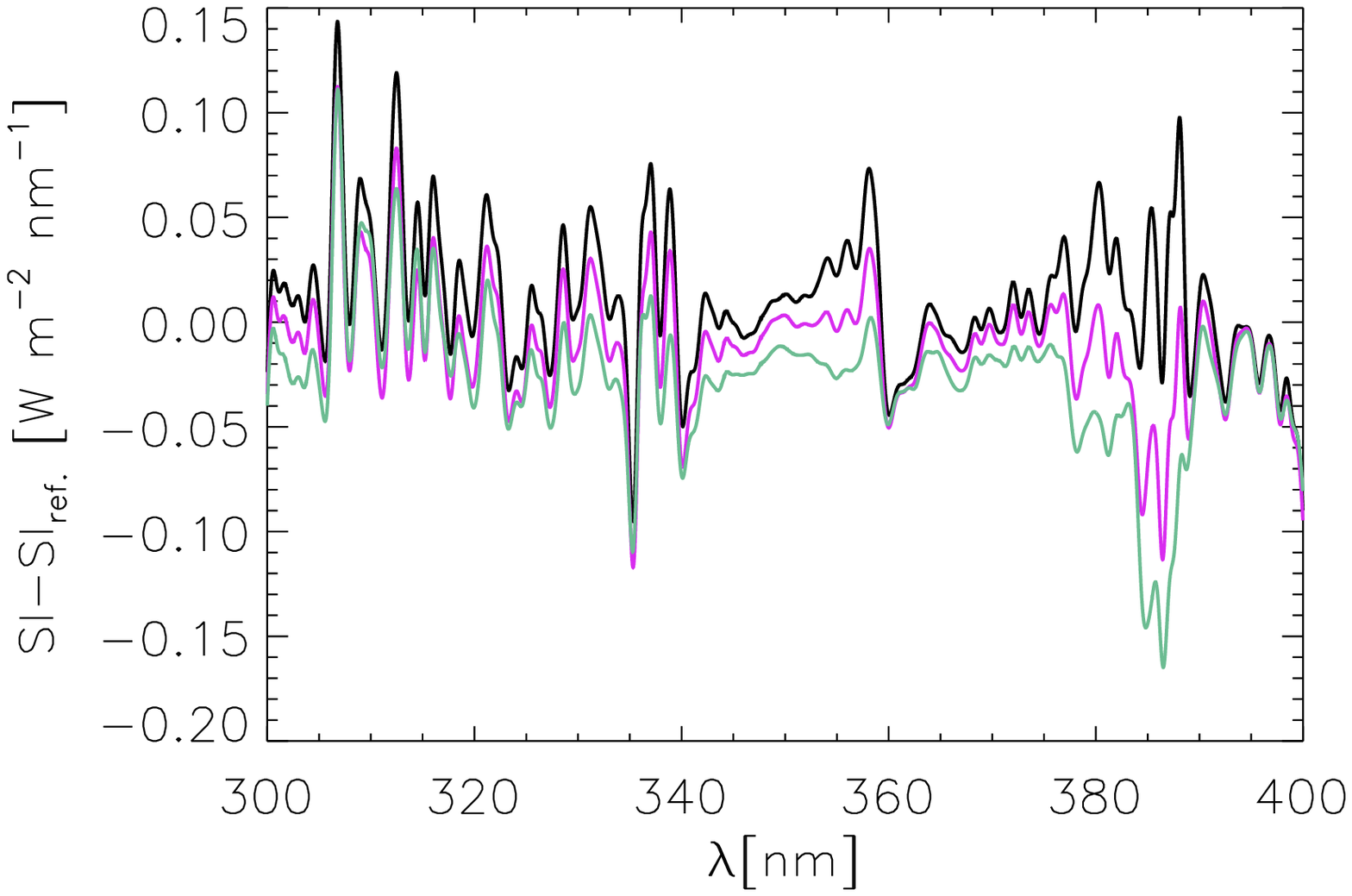}\\
\includegraphics[width=8.5cm]{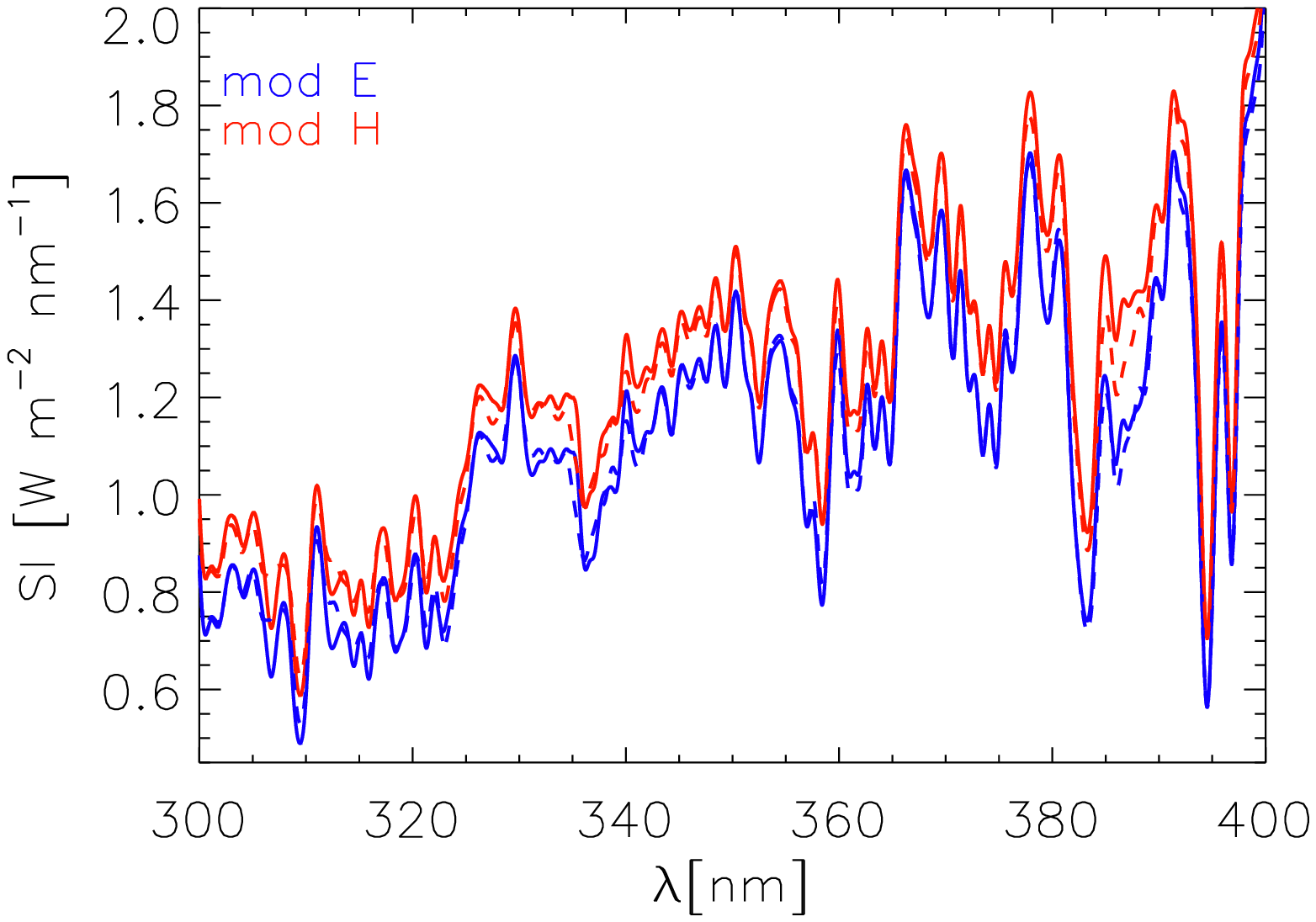} \includegraphics[width=8.5cm]{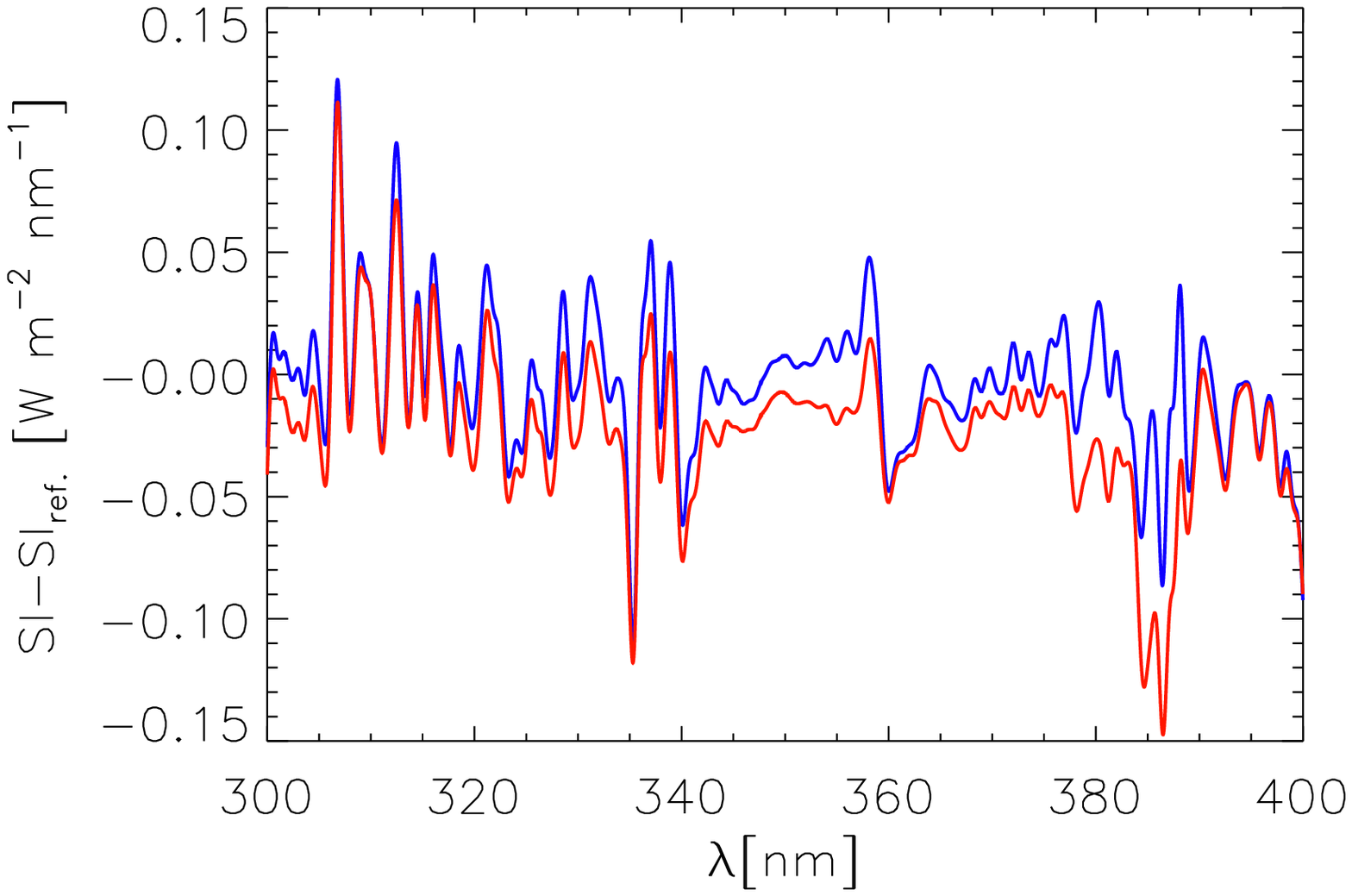}\\
\includegraphics[width=8.5cm]{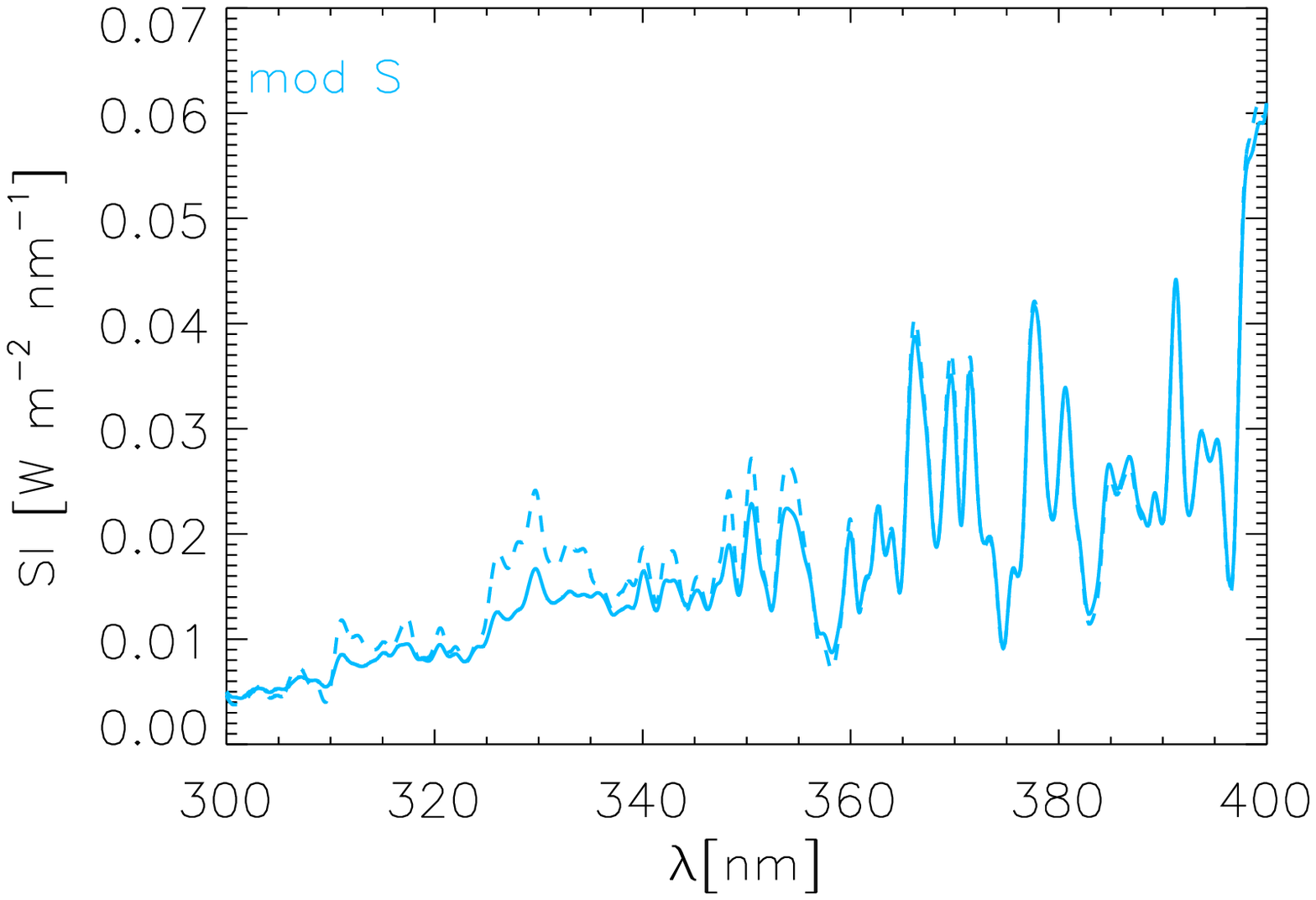} \includegraphics[width=8.5cm]{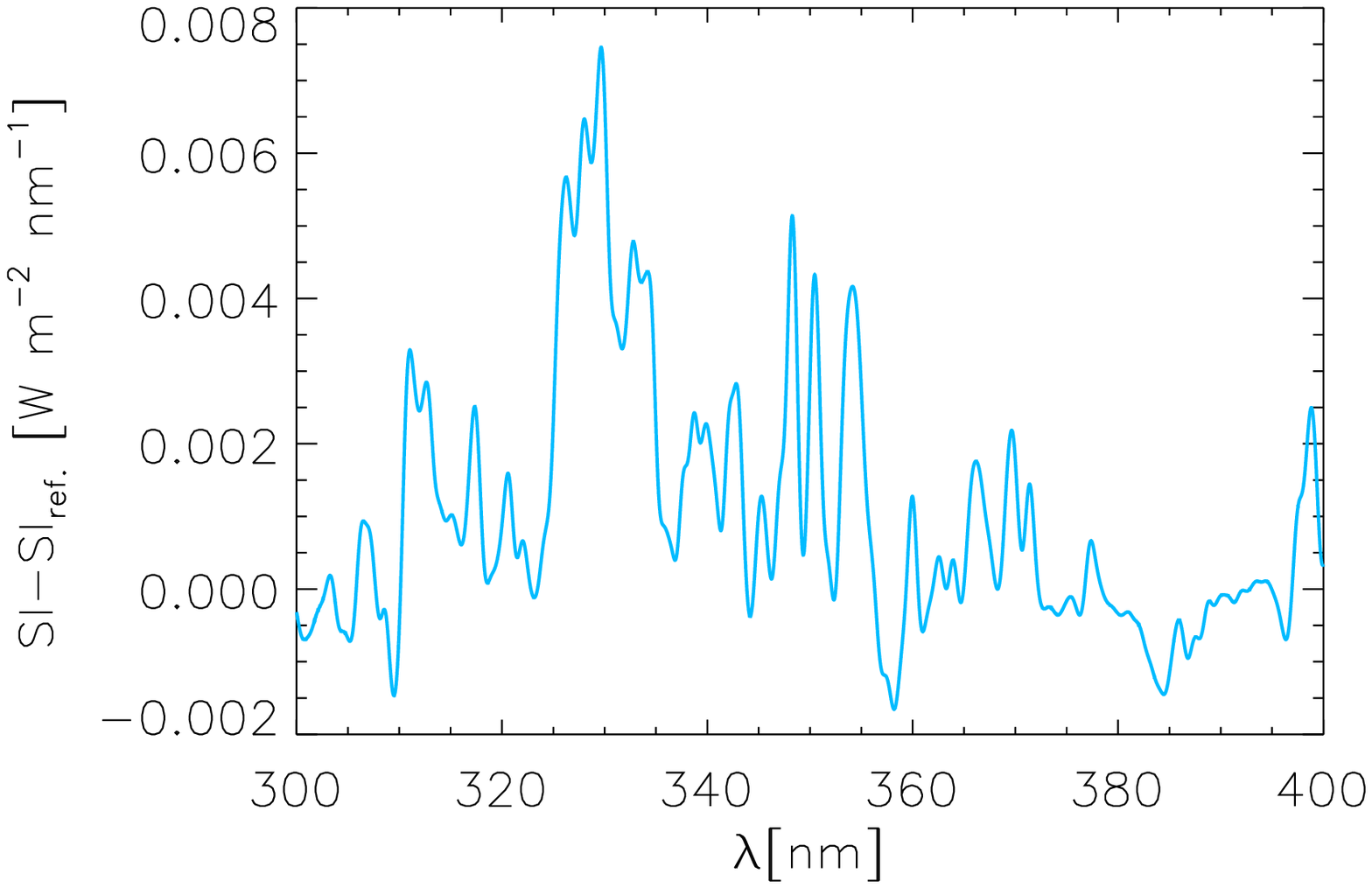}\\
\caption{\label{Fig_300_400} Same as Fig.~\ref{Fig_200_300}, but for the 300-400 nm spectral range.}
\end{figure}

Figure~\ref{Fig_300_400} compares the fluxes converted into irradiance values obtained with the reference synthesis, that is including molecules, and with the synthesis performed without including molecules and applying 
fudge factors to the continuum opacity.
As noticed for the range [230-300] nm, fudging the continuum opacity causes in general underestimate of the radiative emission in ``hot'' models, that is facula and plage models, while produces a slight overestimate 
of the radiative emission in ``cold'' models, such as in the supergranule cell interior and sunspot  models. 

\section {Effects of fudging on estimates of UV Solar Spectral Irradiance variability} \label{sec:variab}

As explained in Sec.~\ref{sec:intro}, fudging the continuum opacity to reproduce the UV spectrum is an approach that has been recently used in radiative transfer codes employed for Solar Irradiance reconstructions. Estimates of radiative properties of magnetic elements are fundamental to understand and model the contribution of such features to solar \citep[e.g.][]{ermolli2011, criscuoli2014, criscuoli2017,yeo2017} and stellar irradiance variability \citep[see e.g.][ for a recent review]{fabbian2017}. 
An indication of the contribution of magnetic structures to irradiance is given by their contrast, that is the ratio between the radiative emission of the magnetic features and the quiet Sun. Photometric contrast of magnetic elements directly estimated from observations is indeed employed in some irradiance reconstruction techniques \citep[e.g.][]{foukal1991,chapman1996,chapman2013} and has often been used to validate atmosphere models of magnetic features \citep[e.g.][]{ermolli2007,ermolli2010}. 

\begin{figure*}
\centering
\includegraphics[width=8.5cm]{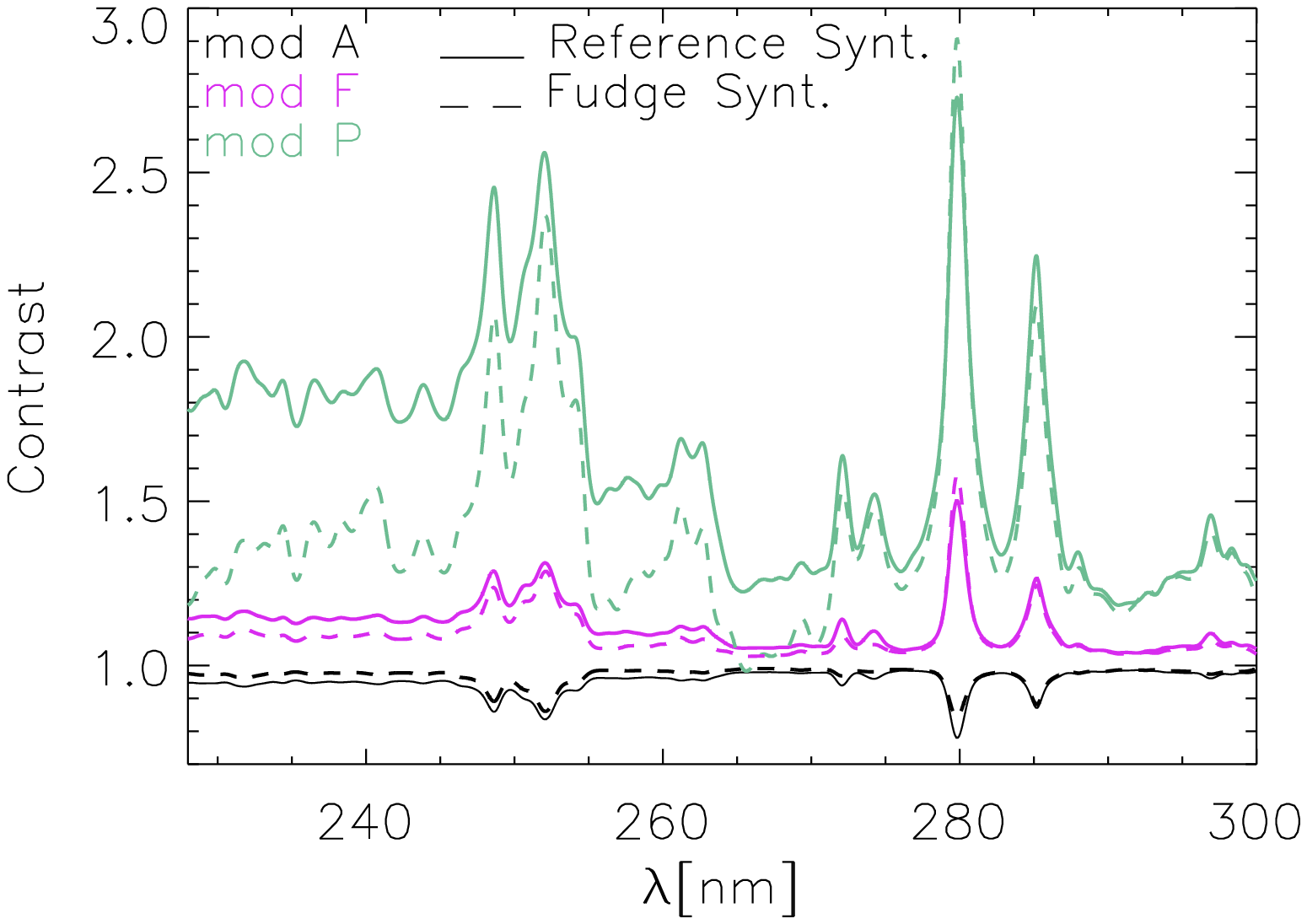}\includegraphics[width=8.5cm]{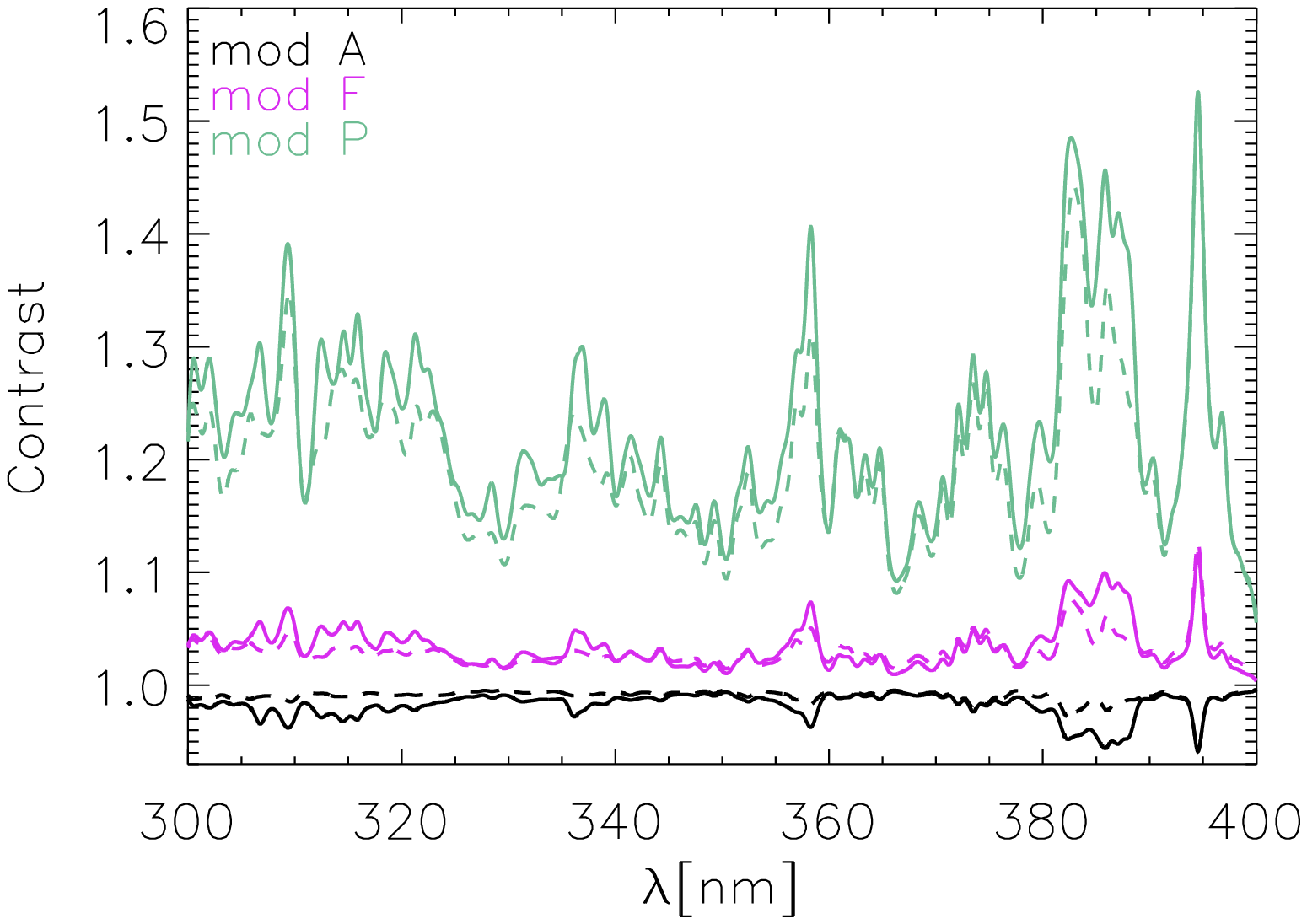}\\
\includegraphics[width=8.5cm]{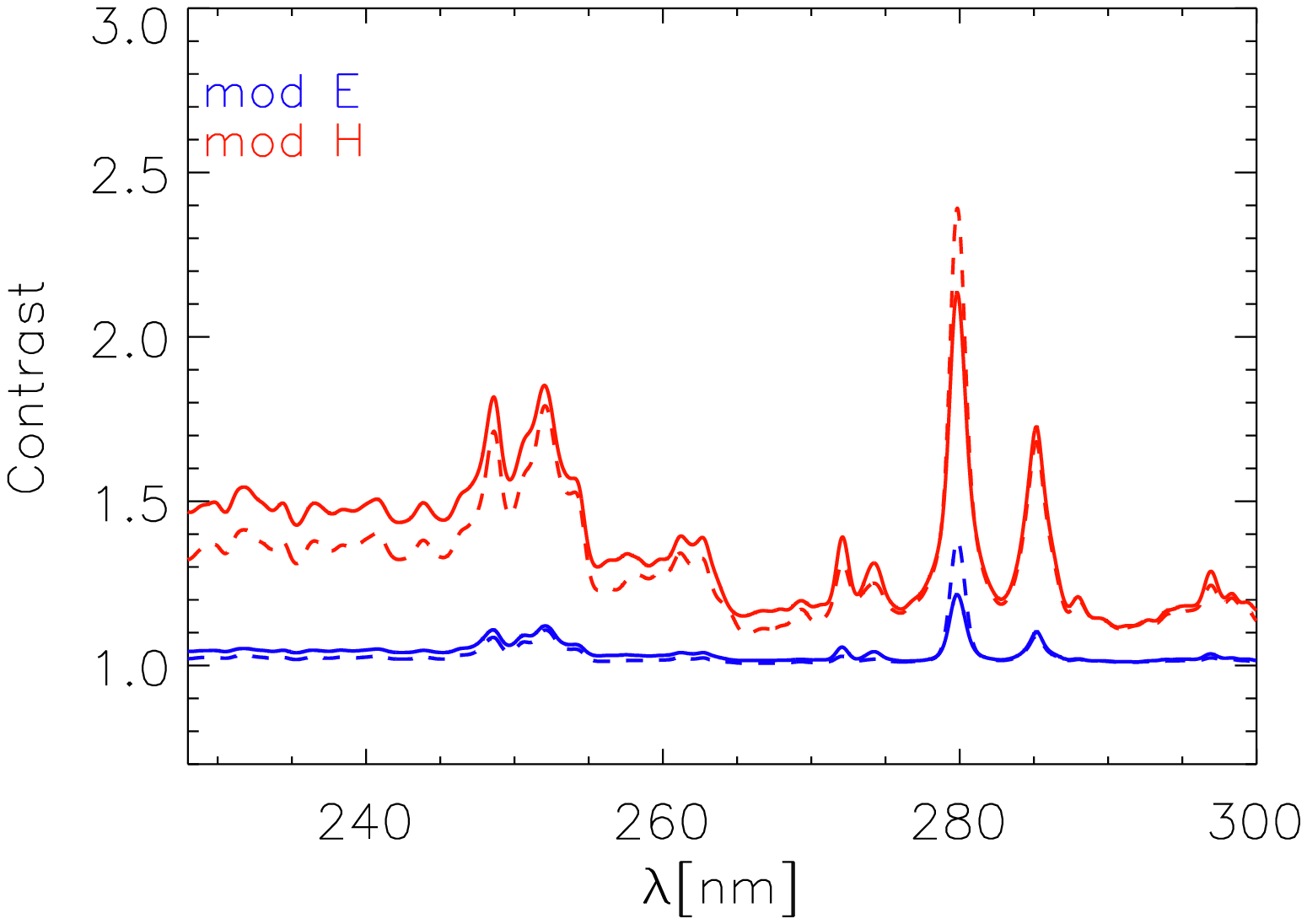}\includegraphics[width=8.5cm]{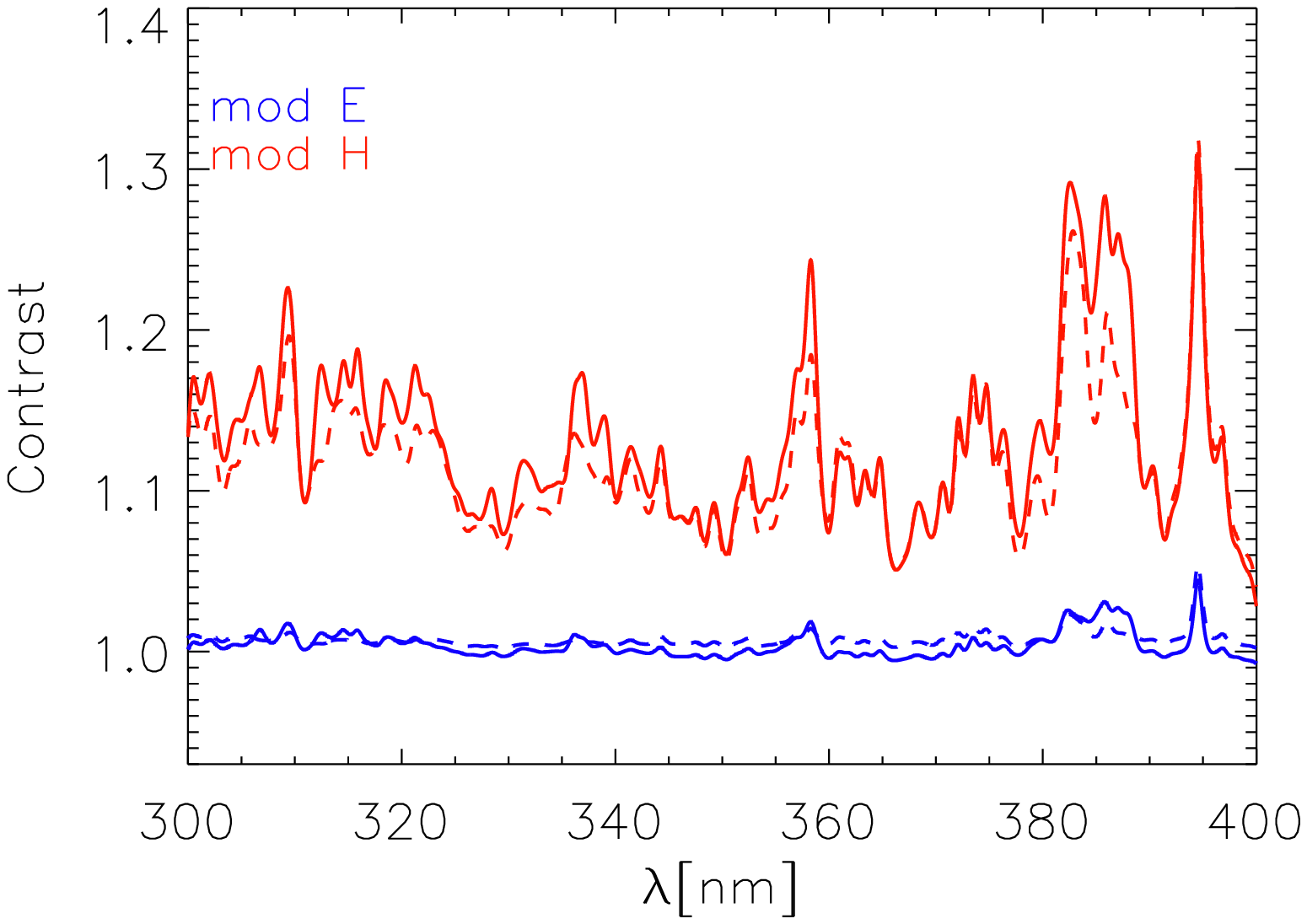}\\
\includegraphics[width=8.5cm]{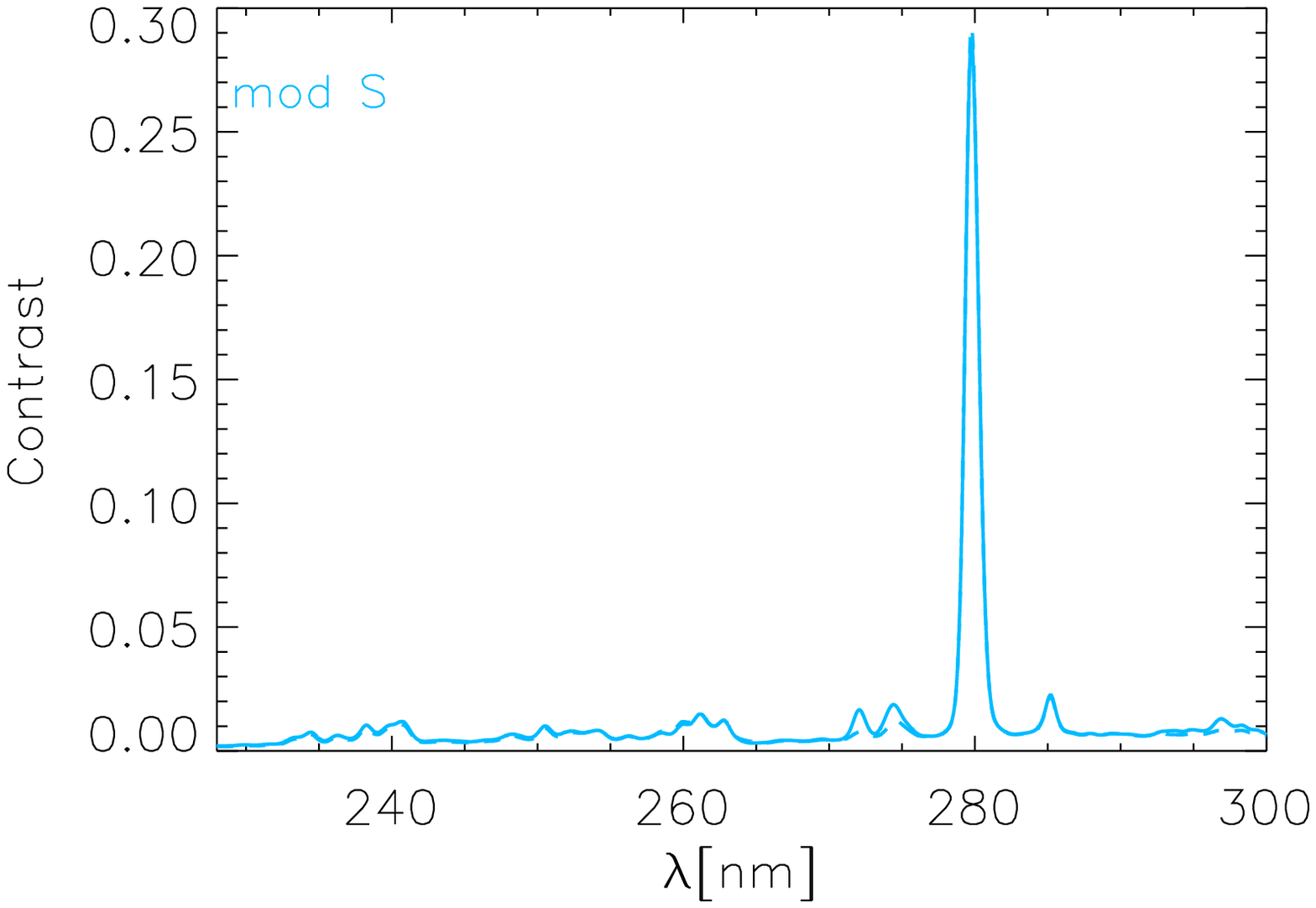}\includegraphics[width=8.5cm]{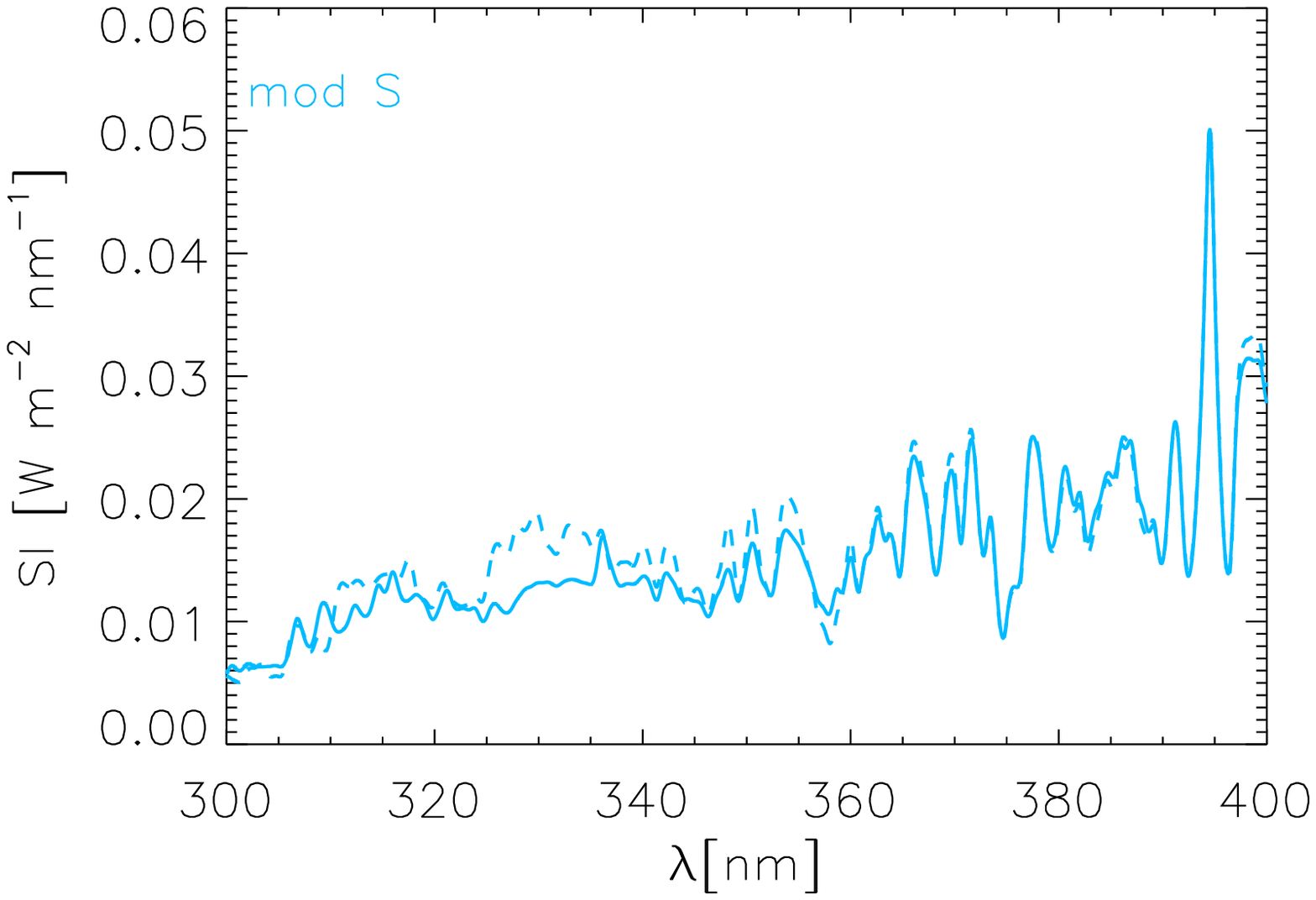}
\caption{\label{Fig_200_400_contradif} Comparison of Spectral Irradiance Contrasts (ratios with respect to quiet Sun model-C).   Continuous: reference synthesis. Dashed: synthesis performed fudging the continuum opacity.
Synthesis performed fudging the opacity typically underestimate
the contrasts.}
\end{figure*}

The contrasts obtained for different magnetic features for the reference synthesis and for the
synthesis performed using the fudge factors described in Sec.~\ref{sec:s200_300} and \ref{sec:s300_400} are shown in Fig.~\ref{Fig_200_400_contradif}. Plots show that,  in spite of the apparent small differences between the spectra shown in 
Fig.~\ref{Fig_200_300} and Fig.~\ref{Fig_300_400}, the difference in contrast between the two synthesis can be up to 0.1, which corresponds
to several tens of a percent of difference. In particular, for models with temperatures higher than the one of model-C, fudging produces less bright magnetic structures, while for models with lower temperature 
fudging produces less dark magnetic structures.

\begin{figure*}
\centering
\includegraphics[width=8.5cm]{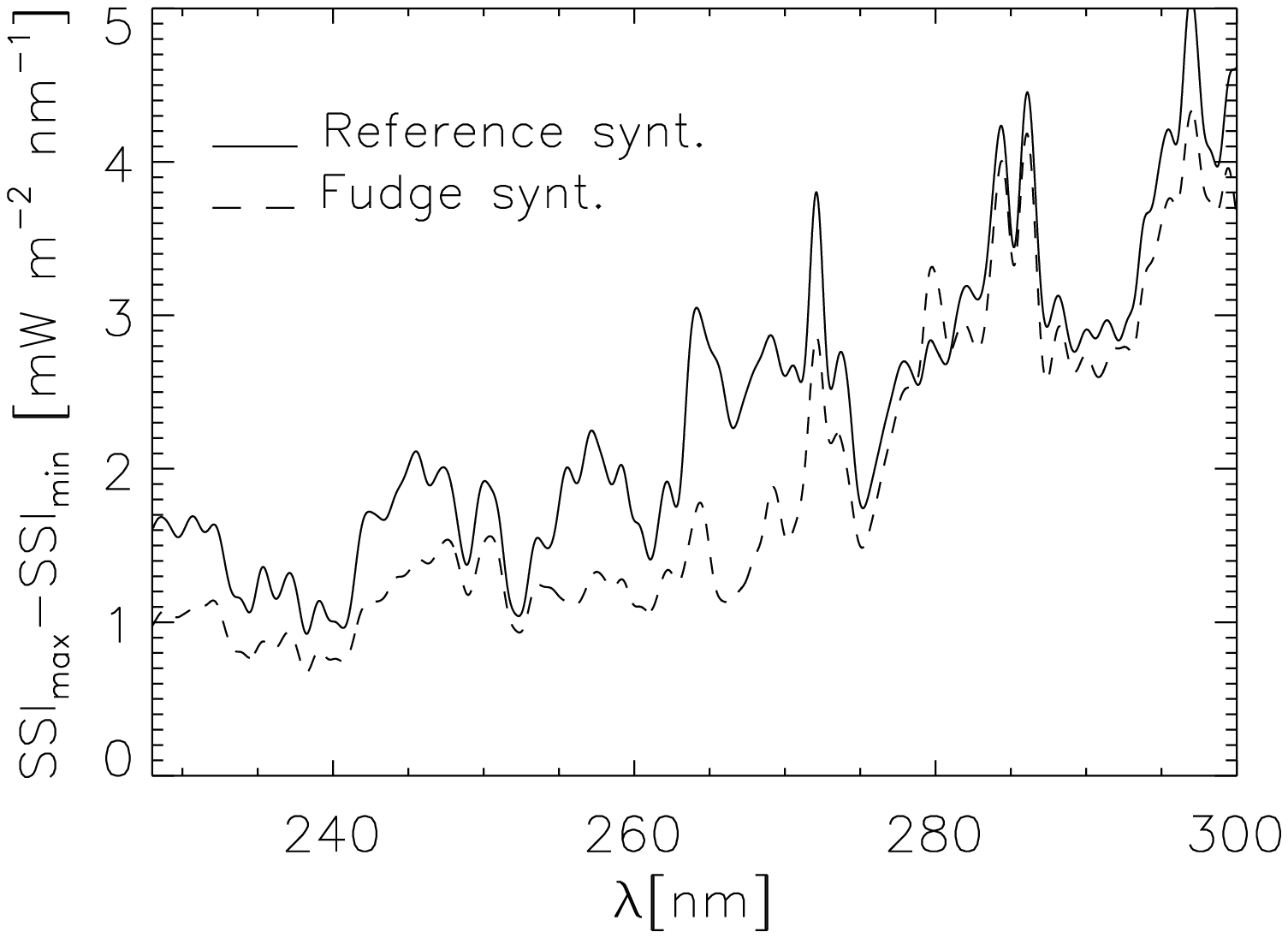}\includegraphics[width=8.5cm]{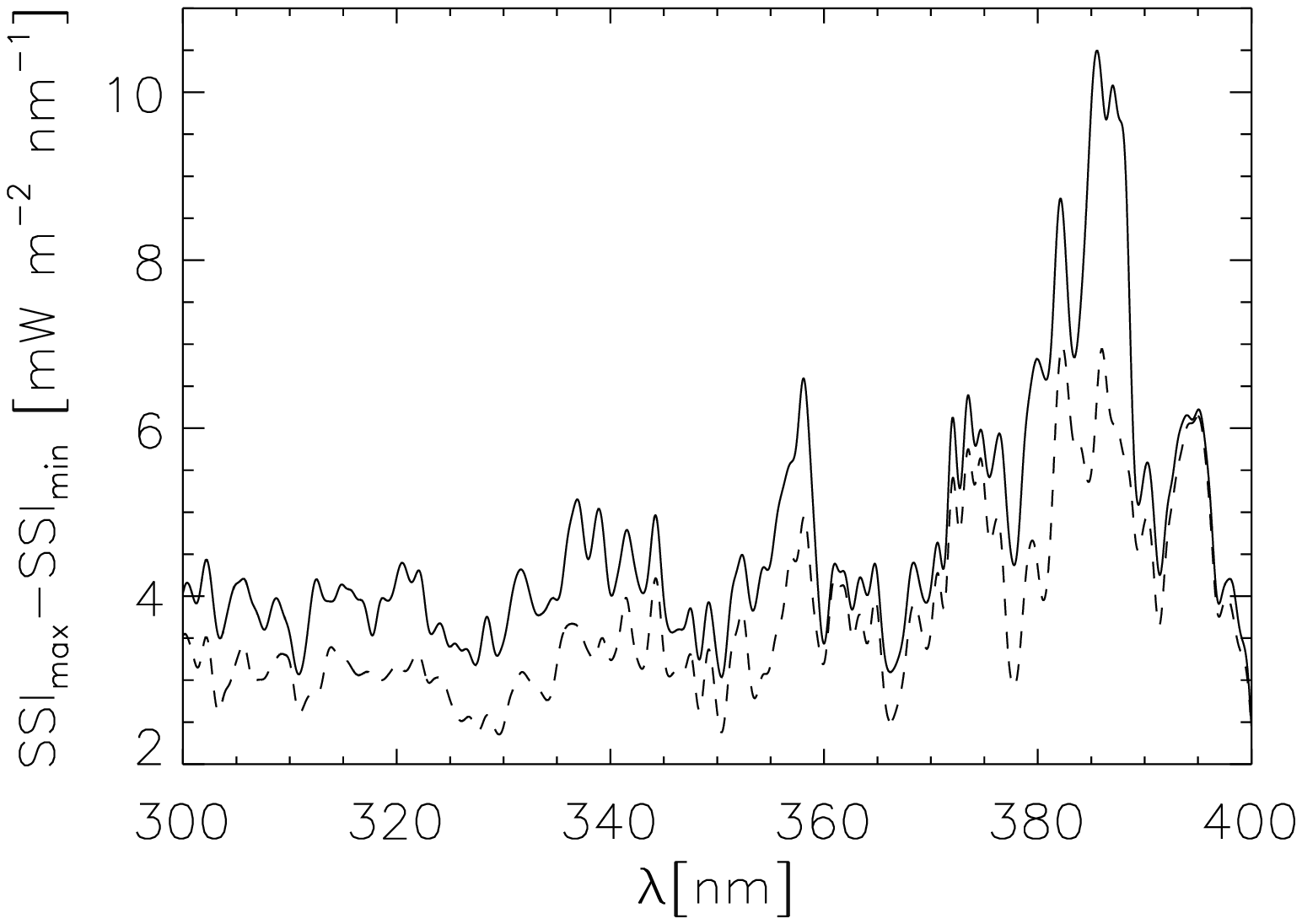}
\caption{\label{Fig_200_400_irradif} SSI variability obtained with the semi-empirical approach described in the text.
Continuous: results obtained from reference synthesis. Dashed: results obtained applying fudge factors to the continuum opacity.  Synthesis performed fudging the continuum opacity systematically underestimate
the variability.}
\end{figure*}

To estimate whether such differences are relevant for reconstructions of solar spectral irradiance variability, 
we employed the semi-empirical approach described in \citet{penza2004,ermolli2011} 
and \citet{criscuoli2018}. As explained in Sec.\ref{sec:intro}, irradiance variability is estimated by weighing the synthetic spectra obtained using atmosphere models representing different types of quiet and active regions, with the area coverage of the corresponding features
as derived by full-disk images. For this analysis, the full-disk data utilized are the Ca II K and red-continuum images acquired with the Precision Solar Photometric Telescope (PSPT) in Hawaii \citep{rast1999}. The dataset and the method employed to identify the magnetic structures are described in details in 
\citet{criscuoli2018}.  
We defined variability as the absolute difference between spectral irradiance computed over a period of minimum (namely the monthly average of April 2009) and maximum (namely the monthly average over August 2012).  The irradiance variability obtained using 
the ``reference'' synthesis and fudging the continuum opacities described in Sec.~\ref{sec:s200_300} and in Sec.~\ref{sec:s300_400} are shown in Fig.~\ref{Fig_200_400_irradif}.

Both plots clearly show that synthesis performed using fudge factors underestimate the variability. In particular, the integrated variability over the range 230-300~nm is 
2.43~$mW m^{-2} nm^{-1}$, 
but it reduces to 1.96 ~$mW m^{-2} nm^{-1}$ 
when applying fudge factors (which corresponds to a derease of about 19\% of the variability). Similarly, the integrated variability over the range 300-400~nm is 4.75~$mW m^{-2} nm^{-1}$
but it reduces to 3.80 ~$mW m^{-2} nm^{-1}$ 
when applying fudge factors (which corresponds to a decrease of the variability of approximately 20\%). Such differences must be ascribed to the overestimate of the continuum opacity found for the brightest magnetic structures (faculae and plage, represented by models - H and -P), shown in Fig.~\ref{Fig_200_300}  and \ref{Fig_300_400}.

\section{Discussion and Conclusions} \label{sec:disc}
\begin{figure*}
\centering
\includegraphics[width=8.5cm]{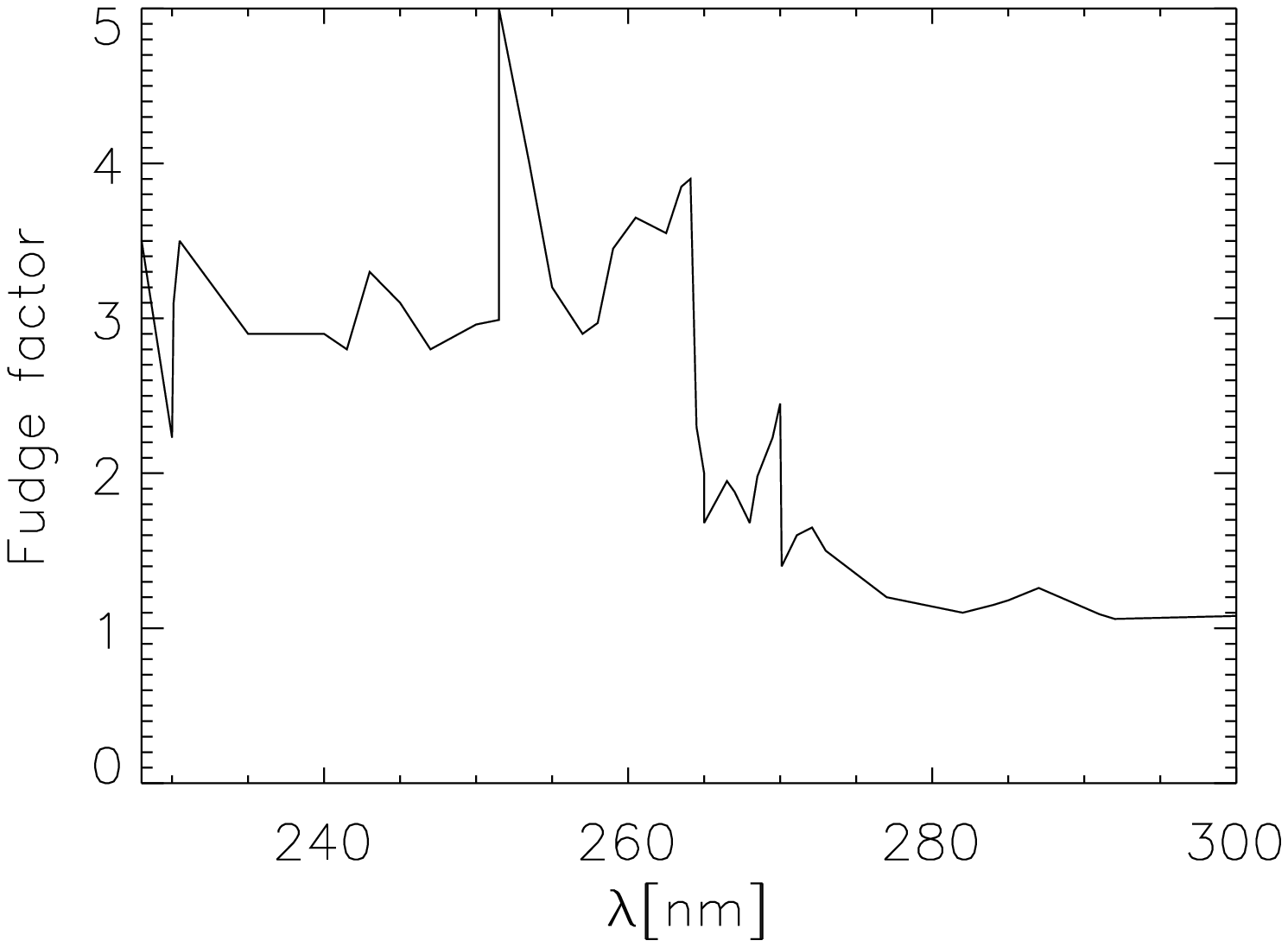} \includegraphics[width=8.5cm]{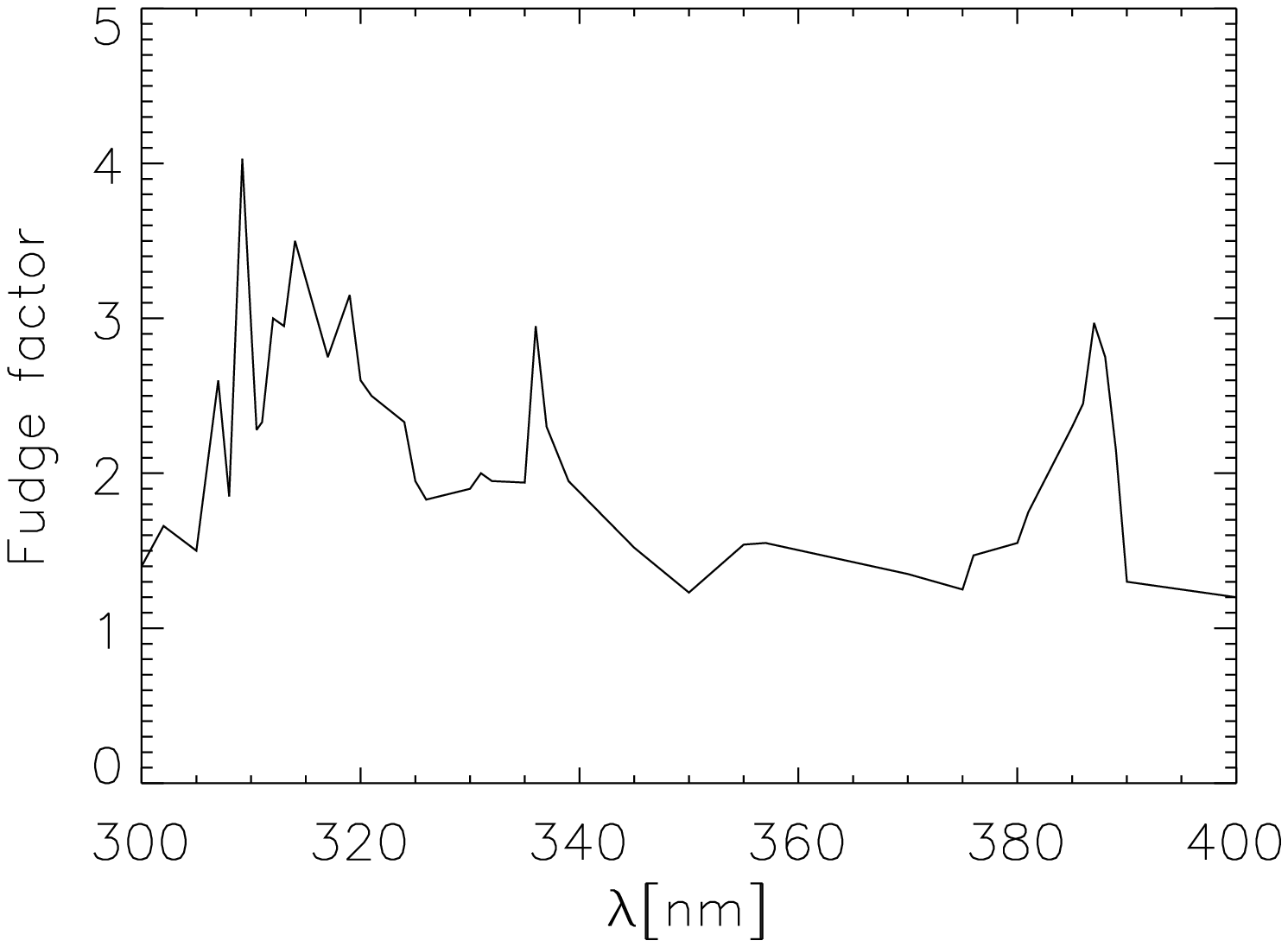}
\caption{\label{Fudge_200_300} Fudge factors in the corresponding spectral ranges. The discontinuity between the values reported in the two ranges results from the different atomic
and molecular inputs employed in the two spectral ranges (see Sec.~\ref{sec:s200_300} and Sec.~\ref{sec:s300_400}). \textbf{As explained in Sec.~\ref{sec:disc}, values in the right panel 
are to be considered upper limits, so that in the region around 300 nm fudge factor values are more likely to be closer to values reported in the left panel.}
}
\end{figure*}

In this paper we have investigated the effects of using continuum fudge factors derived for a quiet Sun model to atmosphere models representing network, faculae and sunspots. Our analysis indicates that in general this approach 
leads to the underestimate
of contrasts and to the consequent underestimate of spectral irradiance variability. It must be stressed that our results are dependent on the atomic and molecular inputs we employed. In particular, for the 230-300 nm range, 
we have compared spectral synthesis computed with Fe I in LTE and in non-LTE assumptions. It is therefore not surprising that the largest differences in contrasts and irradiance variability are obtained at wavelengths shorter than about 275 nm, 
with peaks between 260 and 270 nm, the contribution of Fe I photo-ionization and lines to the total opacity being substantial at these spectral ranges \citep{allendeprieto2003}. Note that the two minima at approximately 265 and 271 nm in Fig.~\ref{Fig_200_300}, visible in all models but model-S, correspond to two Fe I photo-ionization peaks. Similarly, in the range 300-400 nm,
for which we compared synthesis performed with and without molecular lines, the largest differences are found in the range 380-390 nm, where the radiative emission is strongly affected by the presence of CN lines. Not surprisingly, 
as shown by results reported in Fig.~\ref{Fudge_200_300}, the spectral regions with the largest contrast and irradiance variability differences correspond also to the spectral regions with high continuum fudge factors values. 

Qualitatively, the fact that the contrasts of magnetic features are systematically under- or over-estimated when fudge factors are applied to the continuum opacity results from the complex dependence of the opacity on the physical properties of the atmosphere.
In the photosphere and chromosphere the opacity increases rapidly with the temperature \citep[e.g.][]{ferguson2005}, in a way that
can be approximately described as a power-law \textbf{ for temperatures larger than approximately 2500 K}. 
This means that, to obtain comparable changes of formation heights, the fudge factors applied to the continuum opacity should be inversely proportional to the temperature of
the model and that, therefore, the application of fudge factors  same as those suitable for a quiet model are too large for hotter models and too low for colder models. 
In this respect it is interesting to note that \citet{busa2001} found fudge factors values to be inversely proportional to stellar effective temperatures. 
The small differences found for the sunspot model with respect to the other models should be ascribed to the low temperature, which reduces the differences in optical depths 
resulting from using or non-using fudge factors, to the fact that the total opacity shows a less steep dependence on temperature for sunspots typical temperatures in the photosphere
and low-chromosphere \citep[c.f.r. Fig.~9 in][]{ferguson2005} and, finally, to the modest temperature gradient of model-S.    
Moreover, usually atomic and molecular lines tend to weaken in network and facular elements, so that, when fudge factors are employed to take into account of 'missing lines', 
the use of factors derived for quiet models necessarily produces an overestimate of the opacity of network and facular models.

Quantitatively, the amount of uncertainty will depend on the inputs and approximations adopted for the spectral synthesis, that is the set of employed atmosphere models, 
abundance values and opacities.
For the specific case of our calculations, we note that 
the fudge factors employed in the range 230-300 nm are comparable with those adopted by \citet{shapiro2010}, thus suggesting that the UV variability in 
this range might be underestimated of about 19\% also in COSI computations. On the other hand, 
our fudge factors in the range 300-400 nm seem to be systematically higher than those employed in \citet{shapiro2010}, 
thus suggesting that our computations overestimate the effects of fudging in this spectral range. Similarly, results in Fig.~\ref{referefluxes} show that our reference syntheses are closer to the observed spectrum rather than to the spectrum 
utilized to derive the fudge factors, thus suggesting that reconstructions performed using fudging the reference syntheses would be less affected by the uncertainties described above. 
As discussed in \citet{shapiro2010} the choice of aboundance values also affects the synthesis and the derived fudge factors. In particular,
the authors found that the best agreement between their synthesis and the observed spectrum was obtained using the set of opacities published \citet{asplund2005} with the exception of 
Carbon and Nitrogen, whose abundances were those reported in \citet{grevesse1991}. As explained in \citet{tagirov2017} this particular choice was motivated by an incorrect 
computation of H$^{-}$ chemical equilibrium which caused an overestimation of the H$^{-}$ opacity at the visible spectral range. Our computations were not affected by this problem, 
but to investigate the effects of abundances on estimates of fudge factor values, we repeated our synthesis using the set of values from \citet{asplund2005}. We found no appreciable difference
between the synthesis performed using the two sets of abundances in the visible, whereas, in agreement with \citet{shapiro2010}, we found that the use of the values in \citet{asplund2005} 
causes a further excess of flux in the UV range investigated. This, in turn, most likely results from the lower abundance values of Fe and Mg in \citet{asplund2005}, which, as also discussed in 
\citet{shapiro2010}, are major sources of opacity especially in the range [200-300] nm. Our analysis therefore confirms that the use of the set of \citet{asplund2005} abundances implies higher 
fudge factor values and would therefore cause a larger underestimate of the variability.

Finally, the choice of the set of atmosphere models also affects the estimates of irradiance variability, not only because  different quiet-Sun models provide different estimates of the UV flux 
(and therefore different estimates of the fudge factors) but also because of the different temperature and density stratification of models representing magnetic and quiet features. 

Comparison of UV irradiance reconstructions obtained with different techniques shows that in general reconstructions making use of non-LTE synthesis produce larger variability 
than reconstructions making use of proxies or LTE synthesis \citep{ermolli2013, thuillier2014}. Our results indicate that these differences might be even larger,
although still not enough to explain the discrepancies between modeled variability and measurements obtained with the Spectral Irradiance Monitor radiometers \citep{harder2005} aboard the Solar Radiation and Climate Experiment (SORCE) during the descending phase of cycle 23 \citep{2009GeoRL..36.7801H}.   
Note, however, that our analysis cannot rule out the possibility that in semi-empirical NLTE reconstructions the use of fudge factors might in part compensate for an overestimation of the UV variability, which, in turn, might result from the particular assumptions and approaches employed  (e.g. atmosphere models, details in the radiative transfer, identification methods and full-disk data employed). The use of fudge factors would in this case reduce the differences between reconstructions performed using NLTE semi-empirical approaches and other techniques. 

Stellar irradiance reconstructions developed so far mostly make use of
photospheric models and 
LTE synthesis, but the same approaches could be used with models extending to higher layers of stellar atmospheres and non-LTE syntheses. In this case, the effects of 
applying fudge factors to synthetic spectra on the reconstucted irradiance will depend on the fundamental parameters of the star. From the discussion above about the temperature dependence we speculate the 
underestimate of facular component to be relevant for stars with effective temperatures larger than approximately 4000 K \citep[see][]{ferguson2005}. For 'cold' stars as M-dwarfs, we expect 
the effect to be negligible for two reasons. First the temperature stratification of such stars resembles the one of sunspots (c.f.r. Fig.~1 in \citet{fontenla2016} and Fig.~\ref{Fig_temps} above) 
for which we showed spectra to be almost insensitive to the use of fudge factors. Second, results obtained from 3D MHD simulations \citep[e.g.][]{beeck2015a,beeck2015b} show that in photospheres of M-dwarfs small-size 
magnetic structures (which are the building-blocks of plage regions) do not typically produce bright elements due to reduced efficiency of lateral heating. For similar reasons, 
we expect the effect of using fudge factors to be larger for stars whose variability is facula-dominated rather than for stars that are sunspot-dominated. 
We expect instead the use of fudge factors to be less dependent on the surface gravity,  the opacity showing usually a dependence close to linear with the density. 
Indeed, we note that \citet{busa2001} found a linear dependence of the fudge factors on the surface gravity of stars. \textbf{Finally, comparison of syntheses performed using different 
sets of solar opacities suggests fudge factors to increase with the decrease of stellar metallicity, with consequent increase of the uncertainties of stellar variability estimates.}

%

\vspace{5mm}

\section{Acknowledgments}
The National Solar Observatory is operated by the Association of Universities for Research in Astronomy, Inc. (AURA) under cooperative agreement with the National Science Foundation. The author is thankful to Dr. C. Peck for providing  useful comments on the manuscript and Dr. J. Harder for providing the PSPT data. The author also acknowledges the fruitful discussions with the members of the \# 335 science team supported by the International Science Institute (ISSI), Bern.

\software{RH \citep{uitenbroek2001}
          }



\bibliography{biblio} %



\appendix
\section{Quiet Sun spectra}
\begin{figure*}
\centering
\includegraphics[width=8.5cm]{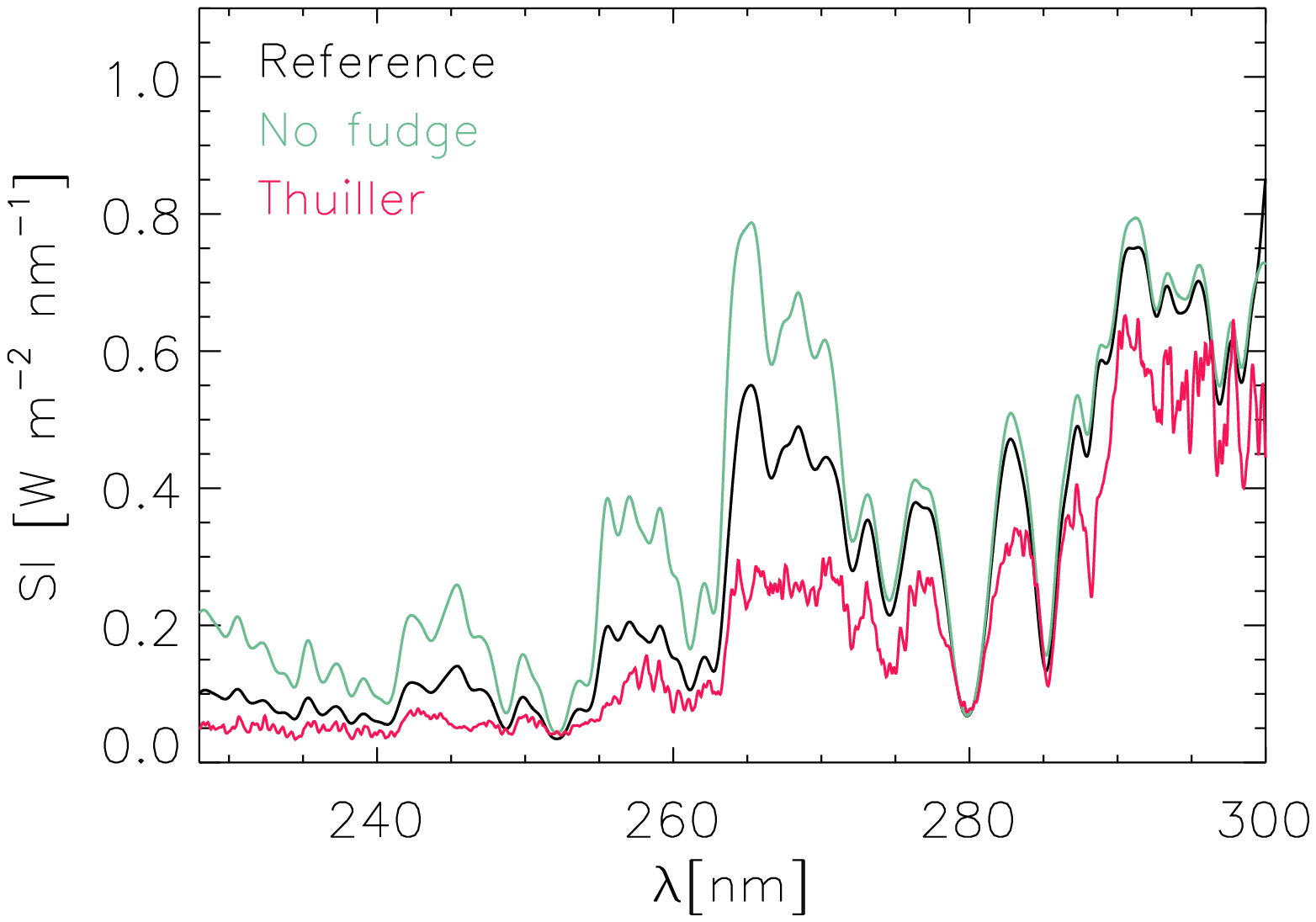} \includegraphics[width=8.5cm]{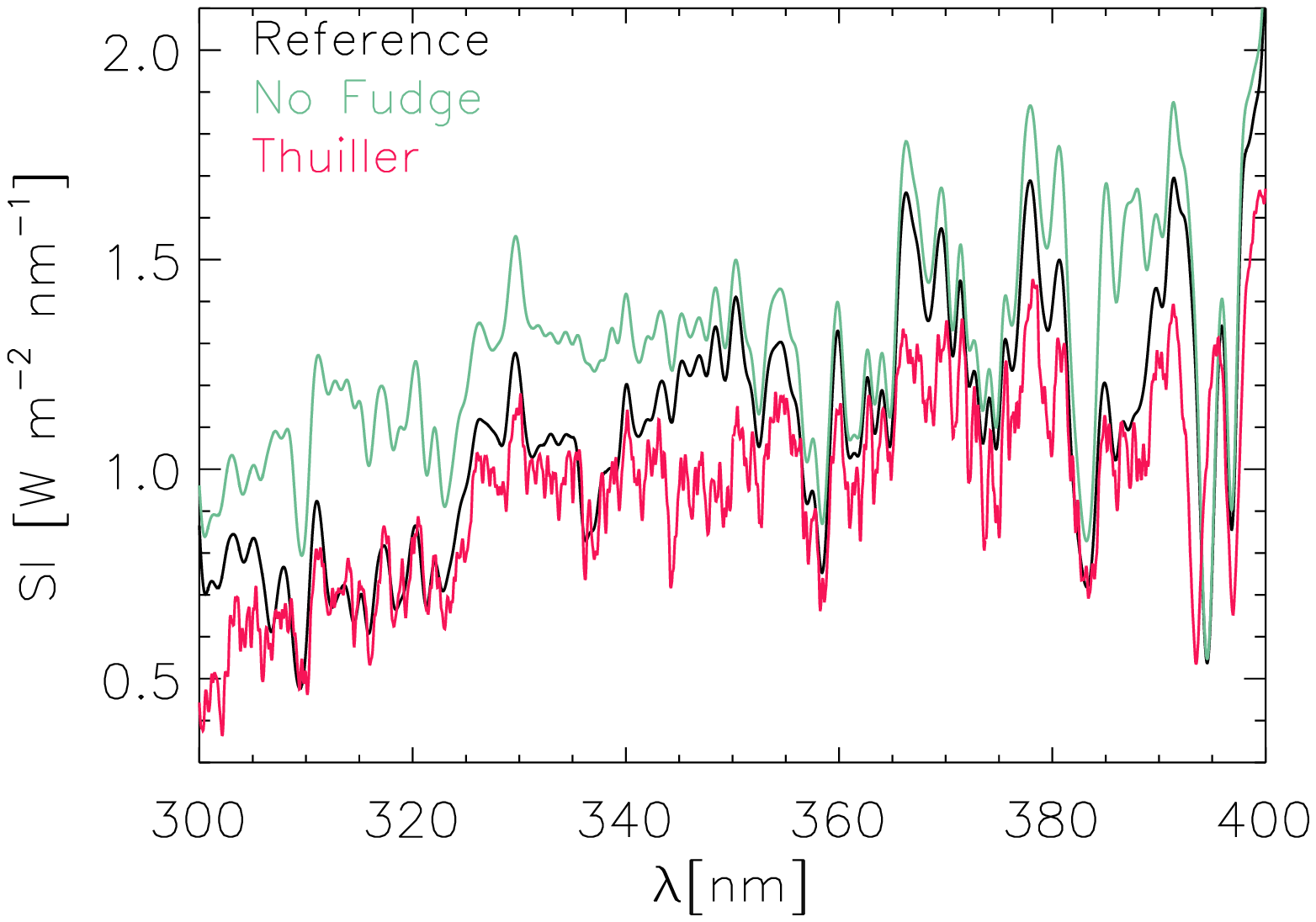}
\caption{\label{referefluxes} Comparison of reference synthesis (black), synthesis performed with simplified set of opacities without applying fudge factors (green) and observed reference
spectrum by \citet[][]{2004AdSpR..34..256T} (red).  }

\end{figure*}

The observed solar spectrum from \citet{2004AdSpR..34..256T} and the synthetic spectra described in Sec.~\ref{sec:s200_300} and Sec.~\ref{sec:s300_400} are illustrated in Fig.~\ref{referefluxes}.  To facilitate the comparison, spectra were convolved with a 1~nm wide Gaussian function.

\end{document}